\documentclass[epj]{svjour}
\usepackage{amsfonts}
\usepackage{latexsym}
\usepackage{graphics}
\usepackage{epsfig,amsmath,graphicx,amssymb,float,,bm}

\begin{document}

\title{Spin squeezing in a bimodal condensate:
spatial dynamics and particle losses}

\author{Yun Li\inst{1,2}, P. Treutlein\inst{3}, J. Reichel\inst{1} \and A. Sinatra\inst{1}}
\institute{Laboratoire Kastler Brossel, ENS, UPMC and CNRS, 24 rue
Lhomond, 75231 Paris Cedex 05, France \and State Key Laboratory of
Precision Spectroscopy, Department of Physics, East China Normal
University, Shanghai 200062, China \and Max-Planck-Institut f\"ur
Quantenoptik and Fakult\"at f\"ur Physik der
Ludwig-Maximilians-Universit\"at, Schellingstrasse 4, 80799
M\"unchen, Germany}
\date{Received: date / Revised version: date}
%
\abstract{We propose an analytical method to study the entangled
spatial and spin dynamics of interacting bimodal Bose-Einstein
condensates. We show that at particular times during the evolution
spatial and spin dynamics disentangle and the spin squeezing can be
predicted by a simple two-mode model.  We calculate the maximum spin
squeezing achievable in experimentally relevant situations with
Sodium or Rubidium bimodal condensates, including the effect of the
dynamics and of one, two and three-body losses.
\PACS{
      {PACS-03.75.Gg}{Entanglement and decoherence in Bose-Einstein condensates}   \and
      {PACS-42.50.Dv}{Quantum state engineering and measurements} \and
      {PACS-03.75.Kk}{Dynamic properties of condensates; collective and hydrodynamic excitations, superfluid flow} \and
      {PACS-03.75.Mn}{Multicomponent condensates; spinor condensates}
     } 
} 

\maketitle

\section{Introduction}
In atomic systems effective spins are collective variables that can
be defined in terms of orthogonal bosonic modes. In this paper the
two modes we consider are two different internal states of the atoms
in a bimodal Bose-Einstein condensate. States with a large first
order coherence between the two modes, that is with a large mean
value of the effective spin component in the equatorial plane of the
Bloch sphere, can still differ by their spin fluctuations. For an
uncorrelated ensemble of atoms, the quantum noise is evenly
distributed among the spin components orthogonal to the mean spin.
However quantum correlations can redistribute this noise and reduce
the variance of one spin quadrature with respect to the uncorrelated
case, achieving spin squeezing \cite{Ueda,manip}. Spin-squeezed
states are multi-particle entangled states that have practical
interest in atom interferometry, and high precision spectroscopy
\cite{Wineland}. Quantum entanglement to improve the precision of
spectroscopic measurements has already been used with trapped ions
\cite{Liebfried} and it could be used in atomic clocks where the
standard quantum limit has already been reached \cite{Santarelli}.

A promising all-atomic route to create spin squeezing in bimodal
condensates, proposed in \cite{Nature}, relies on the Kerr-type non
linearity due to elastic interactions between atoms. Quite
analogously to what happens to a coherent state in a nonlinear Kerr
medium in optics \cite{coherent}, an initial ``phase state'' or
coherent spin state, where all the effective spins point at the same
direction, dynamically evolves into a correlated spin-squeezed
state. A straightforward way to produce the initial phase state in a
bimodal condensate is to start with one atomic condensate in a given
internal state $a$ and perform a $\pi/2$-pulse coupling coherently
the internal state $a$ to a second internal state $b$
\cite{Cornell}. However, as the strength of the interactions between
two atoms $a-a$, $b-b$ and $a-b$ are in general different, the
change in the mean field energy excites the spatial dynamics of the
condensate wave functions. In the evolution subsequent to the pulse,
the spin dynamics creating squeezing and the spatial dynamics are
entangled \cite{Nature,EPJD,Sorensen,Dutton} and occur on the same
time scale set by an effective interaction parameter $\chi$. This
makes it {\it a priori} more difficult to obtain simple analytical
results.

In this paper we develop  a simple formalism which allows us to
calculate analytically or semi analytically the effect of the
spatial dynamics on spin squeezing. In Section \ref{sec:dyn} we
present our dynamic model. Using our treatment we show that at
particular times in the evolution the spatial dynamics and the spin
dynamics disentangle and the dynamical model gives the same results
as a simple two-mode model. We also identify configurations of
parameters in which the simple two mode-model is a good
approximation at all times. Restricting to a two-mode model, in
Section \ref{sec:twomodes} we generalize our analytical results of
\cite{PRLlosses} on optimal spin squeezing in presence of particle
losses to the case of overlapping and non-symmetric condensates.

In Sections \ref{sec:results_over} and  \ref{sec:Galette}, we apply
our treatment to cases of practical interest. We first consider a
bimodal ${}^{87}$Rb condensate. Rb is one of the most common atoms
in BEC experiments and it is a good candidate for atomic clocks
using trapped atoms on a chip \cite{clockRb}. Restricting to states
which are equally affected by a magnetic field to first order, the
most common choices are $|F=1, m=-1\rangle$ and $|F=2, m=1\rangle$
which can be magnetically trapped, or $|F=1, m=1\rangle$ and $|F=2,
m=-1\rangle$ that must be trapped optically but for which there
exists a low-field Feshbach resonance which can be used to reduce
the inter-species scattering length \cite{Sengstock,Widera}. Indeed
a particular feature of these Rb states is that the three $s$-wave
scattering lengths characterizing interactions between $a-a$, $b-b$
and $a-b$ atoms are very close to each other. A consequence is that
the squeezing dynamics is very slow when the two condensates
overlap. The inter-species Feshbach resonance can be used to
overcome this problem and speed up the dynamics  \cite{Widera}.

In schemes involving the $|F=2, m=\pm1\rangle$ of Rubidium, the main
limit to the maximum squeezing achievable is set by the large
two-body losses rate in these states. As a second case of
experimental interest we then consider Na atoms in the $|F=1,m_F=\pm
1\rangle$ states \cite{Nature}. Although theses states have opposite
shifts in a magnetic field, they present the advantage of negligible
two-body losses. Using our analytical optimization procedure, we
calculate the maximum squeezing achievable in this system including
the effect of spatial dynamics and particle losses.

In Section \ref{sec:Galette} we examine a different scenario for Rb
condensates in which, instead of changing the scattering length, one
would spatially separate the two condensates after the mixing
$\pi/2$ pulse and hold them separately during a well chosen
squeezing time. An interesting feature of this scheme is that the
squeezing dynamics acts only when the clouds are spatially separated
and it freezes out when the two clouds are put back together so that
one could prepare a spin squeezed state and then keep it for a
certain time \cite{PRLlosses}. State-selective potentials for
${}^{87}$Rb in $|F=1, m=-1\rangle$ and $|F=2, m=1\rangle$
\cite{clockRb} have recently been implemented on an atom chip, and
such scheme could be of experimental interest.

\section{Dynamical spin squeezing model}
\label{sec:dyn}
In this section we develop and compare dynamical models for spin
squeezing. No losses will be taken into account in this section.

\subsection{State evolution}
We consider the model Hamiltonian
\begin{eqnarray}
H&=&\int d^3\vec{r} \: \sum_{\varepsilon=a,b} \left[
\hat{\psi}^\dagger_\varepsilon h_\varepsilon \hat{\psi}_\varepsilon
+ \frac{1}{2} g_{\varepsilon \varepsilon}
\hat{\psi}^\dagger_\varepsilon \hat{\psi}^\dagger_\varepsilon
\hat{\psi}_\varepsilon \hat{\psi}_\varepsilon \right]
\nonumber \\
&+&  g_{ab} \hat{\psi}^\dagger_b \hat{\psi}^\dagger_a
\hat{\psi}_a \hat{\psi}_b \,
\end{eqnarray}
where $h_{\varepsilon}$ is the one-body hamiltonian including
kinetic energy and external trapping potential
\begin{equation}
h_{\varepsilon}= -\frac{\hbar^2\Delta}{2m}+U^{\mbox{
\small ext}}_\varepsilon(\vec{r}\,) \,.
\end{equation}
The interactions constants $g_{\varepsilon \varepsilon'}$ are
related to the corresponding $s$-wave scattering lengths
$g_{\varepsilon \varepsilon'}=4\pi \hbar^2 a_{\varepsilon
\varepsilon'}/M$ characterizing a cold collision between an atom in
state $\varepsilon$ with an atom in state $\varepsilon'$
($\varepsilon, \varepsilon'=a,b$), and $M$ is the mass of one atom.

We assume that we start from a condensate with $N$ atoms in the
internal state $a$; the stationary wave function of the condensate
is $\phi_0(\vec{r}\,)$. After a $\pi/2$ pulse, a phase state is
created, which is our initial state:
\begin{equation}
|\Psi(0)\rangle=\frac{1}{\sqrt{N!}}\left[ C_a
a^\dag_{|\phi_0\rangle} + C_b b^\dag_{|\phi_0\rangle}
\right]^{N} |0\rangle \label{eq:psi0}
\end{equation}
where $C_a$, $C_b$ are mixing coefficients with $|C_a|^2+|C_b|^2=1$
and the operator $a^\dag_{|\phi_0\rangle}$ creates a particle in the
internal state $a$ with wave function $\phi_0$. To describe the
entangled evolution of the spin dynamics and the external dynamics
of the wave functions, it is convenient to introduce Fock states
with well defined number of particles in $|a\rangle$ and
$|b\rangle$, these numbers being preserved during time evolution
subsequent to the mixing pulse. Expanded over the Fock states, the
initial state (\ref{eq:psi0}) reads:
\begin{equation}\label{eq:initial}
|\Psi(0)\rangle=\sum^N_{N_a=0}\left(\frac{N!}{N_a!N_b!}\right
)^{1/2}C_a^{N_a}C_b^{N_b}|N_a:\phi_0,N_b:\phi_0\rangle,
\end{equation}
where $N_b=N-N_a$, and
\begin{equation}
|N_a:\phi_a,N_b:\phi_b\rangle=\frac{\left[
{a}^\dag_{|\phi_a(N_a,N_b) \rangle}\right]^{N_a}}{\sqrt{N_a!}}
\frac{\left[{b}^\dag_{|\phi_b(N_a,N_b)
\rangle}\right]^{N_b}}{\sqrt{N_b!}} |0\rangle \,. \label{eq:Fock}
\end{equation}

Within an Hartee-Fock type ansatz for the $N$-body state vector, we
calculate the evolution of each Fock state in (\ref{eq:initial}). We
get \cite{EPJD}:
\begin{eqnarray}
&&|N_a:\phi_0,N_b:\phi_0\rangle\rightarrow
e^{-iA(N_a,N_b;\,t)/\hbar}  \nonumber \\
&&\hspace{1cm} \times
|N_a:\phi_a(N_a,N_b;t),N_b:\phi_b(N_a,N_b;t)\rangle \,,
\label{eq:evol_Fock}
\end{eqnarray}
where $\phi_a(N_a,N_b;t)$ and $\phi_b(N_a,N_b;t)$ are solutions of
the coupled Gross-Pitaevskii equations:
\begin{equation}
i\hbar \partial_t \phi_\varepsilon= \left[ h_{\varepsilon} +
(N_\varepsilon-1)g_{\varepsilon\varepsilon}|\phi_\varepsilon|^2
+N_\varepsilon' g_{\varepsilon\varepsilon'}|\phi_{\varepsilon'}|^2
\right ]\phi_\varepsilon \label{eq:GPE}
\end{equation}
here with the initial conditions
\begin{equation}\label{ini_cond}
 \phi_a(0)=\phi_b(0)=\phi_0 \,,
\end{equation}
and the time dependent phase factor $A$ solves:
\begin{eqnarray}
\label{eq:A(t)} \frac{d}{dt}A(N_a,N_b;t)&=&-\sum_{\varepsilon=a,b}
N_\varepsilon(N_\varepsilon-1)\frac{g_{\varepsilon
\varepsilon}}{2}\int d^3r |\phi_\varepsilon|^4 \nonumber \\
&-& N_a N_b g_{ab}\int d^3r |\phi_a|^2|\phi_b|^2 \,.
\end{eqnarray}
With this treatment we fully include the quantum dynamics of the two
condensate modes $a$ and $b$, as one does for the simple two modes
model, but also including the spatial dynamics of the two modes and
their dependence on the number of particles. The approximation we
make is to neglect all the other modes orthogonal to the condensates
which would be populated thermally. An alternative method is to use
a number conserving Bogoliubov theory that explicitly includes the
operators of the condensates as in \cite{Sorensen}. In that case all
the modes are present but the modes orthogonal to the condensates
are treated in a linearized way. In \cite{Sorensen}, the author
compares the number conserving Bogoliubov approach to our approach
using many Gross-Piaevskii equations, also used in \cite{Nature},
and he finds very similar result for the spin squeezing. He also
finds that within the Bogoliuobov approximation the thermally
excited modes strictly do not affect the squeezing in the scheme we
consider here. If the number conserving Bogoliubov has the advantage
of being systematic, our approach, supplemented with a further
approximation (the modulus-phase approximation introduced in Sect.
\ref{sub:modulusphase}) allows us to get some insight and obtain
simple analytical results.

\subsection{Calculation of spin squeezing}

\begin{figure}
\centering
\includegraphics[scale=0.6]{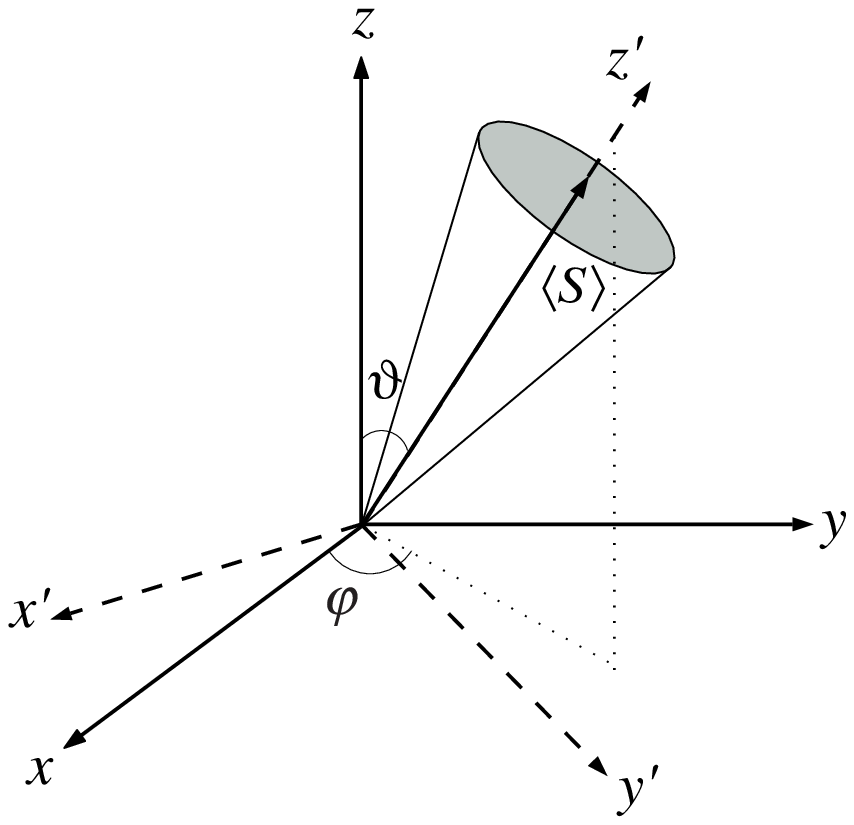}
\includegraphics[scale=0.5]{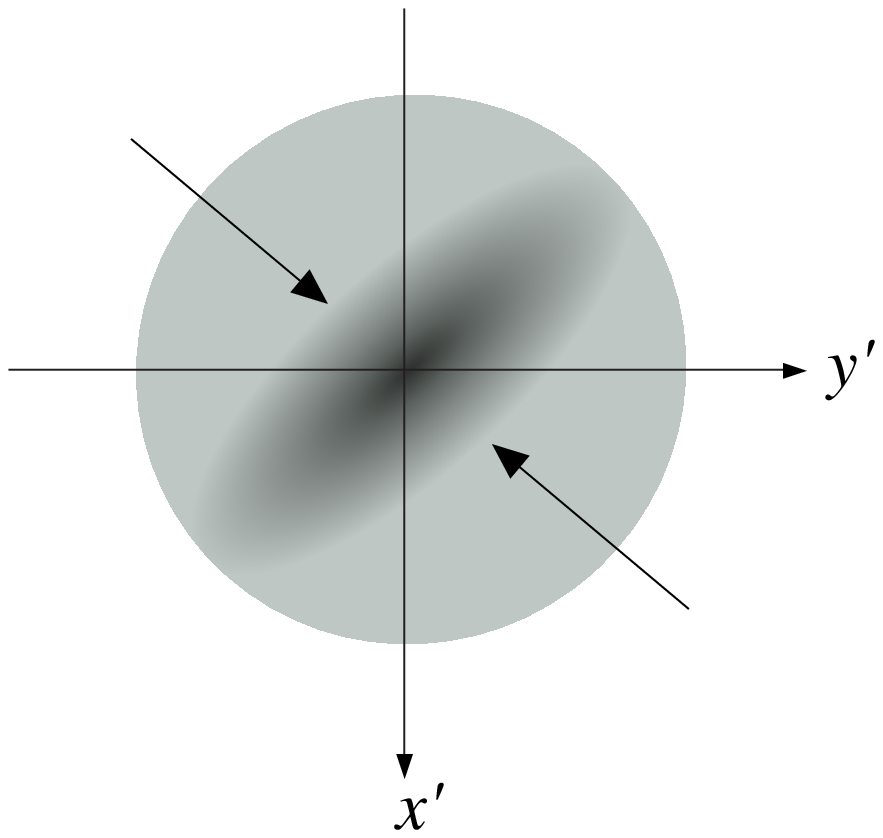}
\caption{(Top) average spin. (Bottom) variance of the the spin
components in the plane orthogonal to the mean spin.} \label{fig1}
\end{figure}

The effective collective spin of a two-components BEC can be
represented on the Bloch sphere as shown in Fig.\ref{fig1}(Top).
Formally, we introduce three spin operators in terms of field
operators \cite{Nature}
\begin{eqnarray}
S_x=&\dfrac{1}{2}\int d^3r[\hat{\psi}^\dag_b
(\vec{r}\,)\hat{\psi}_a(\vec{r}\,)+\hat{\psi}^\dag_a
(\vec{r}\,)\hat{\psi}_b(\vec{r}\,)], \label{eq:Sx} \\
S_y=&\dfrac{i}{2}\int d^3r[\hat{\psi}^\dag_b (\vec{r}\,)\hat{\psi}_a
(\vec{r}\,) -\hat{\psi}^\dag_a
(\vec{r}\,)\hat{\psi}_b(\vec{r}\,)], \label{eq:Sy}\\
S_z=&\dfrac{1}{2}\int d^3r[\hat{\psi}^\dag_a (\vec{r}\,)\hat{\psi}_a
(\vec{r}\,)-\hat{\psi}^\dag_b (\vec{r})\hat{\psi}_b(\vec{r})].
\label{eq:Sz}
\end{eqnarray}
Definitions (\ref{eq:Sx})-(\ref{eq:Sz}) explicitly take into account
the spatial wave functions of the condensate and depend in
particular on the overlap between the two modes.

Referring to the Fig.\ref{fig1}(Top) we introduce the polar angles
$\vartheta$ and $\varphi$ giving the direction $z'$ of the mean
spin; $\vartheta$ determines the relative mean atom number in the
two internal states, $\cos \vartheta=|C_a|^2-|C_b|^2$, while the
azimuthal angle $\varphi$ corresponds to the relative phase between
the components, $\varphi=\arg(C_a^\ast C_b)$.

The minimal variance of the spin in the plane $(x',y')$ orthogonal
to the mean spin $\Delta S_{\perp,\mbox{\small min}}^2$, represented
in Fig. \ref{fig1}(Bottom), is given by
\begin{equation}
\Delta S_{\perp,\mbox{\small min}}^2=\frac{1}{2} (\Delta
S_{x'}^2+\Delta S_{y'}^2 - |\Delta S_-^2|)
\end{equation}
where we introduced
\begin{equation}
S_{-}=S_{x'}-i S_{y'} \,.
\end{equation}

The degree of squeezing is then quantified by the parameter
\cite{Nature,Wineland}
\begin{equation}
\xi^2=\frac{N \Delta S_{\perp,\mbox{\small min}}^2}{\langle S
\rangle^2 }, \label{eq:xi2gen}
\end{equation}
where $\langle S \rangle$ is the length of the average spin.

When expressed in the original frame of reference, the minimal
variance in the orthogonal plane is:
\begin{eqnarray}
\label{eq:squeez} \Delta S_{\perp,\mbox{\small min}}^2&=&
\frac{1}{2}(\cos^2\vartheta\cos^2 \varphi+\sin^2\varphi) \Delta S_x^2 \nonumber \\
&+&\frac{1}{2}(\cos^2\vartheta\sin^2\varphi+\cos^2\varphi) \Delta S_y^2 \nonumber \\
&+&\frac{1}{2}\sin^2\vartheta\Delta S_z^2
-\frac{1}{4}\sin^2\vartheta\sin
2\varphi\Delta_{xy}\nonumber\\
&-&\frac{1}{4}\sin 2\vartheta\cos\varphi \Delta_{zx}
-\frac{1}{4}\sin 2\vartheta\sin\varphi
\Delta_{yz}\nonumber \\
& -& \frac{1}{2}\sqrt{\tilde{A}^2+\tilde{B}^2}
\end{eqnarray}
where
\begin{eqnarray}
\tilde{A}&=&(\sin^2\varphi-\cos^2\vartheta\cos^2\varphi)\Delta S_x^2
\phantom{\frac{1}{2}}\nonumber \\
&+&(\cos^2\varphi-\cos^2\vartheta\sin^2\varphi)\Delta S_y^2
-\sin^2\vartheta\Delta S_z^2 \nonumber \\
&-&\frac{1}{2}(1+\cos^2\vartheta)\sin 2\varphi \Delta_{xy}+
\frac{1}{2}
\sin 2\vartheta\cos\varphi\Delta_{zx} \nonumber \\
&+& \frac{1}{2}\sin 2\vartheta\sin\varphi \Delta_{yz} \:; \\
\tilde{B}&=& \cos\vartheta\sin
2\varphi(\Delta S_x^2-\Delta S_y^2)-\cos\vartheta
\cos 2\varphi\Delta_{xy} \nonumber \\
&-&\sin\vartheta\sin\varphi\Delta_{zx}+\sin\vartheta\cos\varphi
\Delta_{yz} \:;
\end{eqnarray}
and where we introduced the correlations
\begin{equation}
\Delta_{ij}=\langle S_i S_j+S_j S_i \rangle-2\langle S_i \rangle
\langle S_j \rangle, \hspace{0.5cm}  i\neq j=x,y,z \,.
\end{equation}
The spin squeezing is then calculated in terms of averages of field
operators products, with the state of the system at time $t$,
obtained by evolving equation (\ref{eq:initial}) with equation
(\ref{eq:evol_Fock}). To calculate the averages one needs to compute
the action of the field operators $\hat{\psi}_a$ $\hat{\psi}_b$ on
the Fock states (\ref{eq:Fock}) \cite{notecalc1},

\begin{eqnarray}
&\hat{\psi}_a(\vec{r})&|N_a:\phi_a(N_a,N_b),N_b:\phi_b(N_a,N_b)
\rangle \nonumber \\& =& \phi_a(N_a,N_b,\vec{r})\sqrt{N_a} \nonumber \\
&\times & |N_a-1:\phi_a(N_a,N_b), N_b:
\phi_b(N_a,N_b) \rangle, \label{field_statea}\\
&\hat{\psi}_b(\vec{r})&|N_a:\phi_a(N_a,N_b),N_b:\phi_b(N_a,N_b)
\rangle \nonumber \\ & = &\phi_b(N_a,N_b,\vec{r})\sqrt{N_b} \nonumber \\
&\times& |N_a:\phi_a(N_a,N_b), N_b-1: \phi_b(N_a,N_b) \rangle. \label{field_stateb}
\end{eqnarray}
The explicit expressions of the averages needed to calculate the
spin squeezing parameter are given in Appendix \ref{app:averages}.
These quantum averages correspond to an initial state with a
well-defined number of particles $N$. In case of fluctuations in the
total number of particles where the density matrix of the system is
a statistical mixture of states with a different number of
particles, a further averaging of $N$ over a probability
distribution $P(N)$ is needed \cite{EPJD,EPJDLoss}.

\subsection{Dynamical modulus-phase approach}
\label{sub:modulusphase}

In principle, equations (\ref{eq:GPE})-(\ref{eq:A(t)}) can be solved
numerically for each Fock state in the sum equation
(\ref{eq:initial}), and the squeezing can be computed as explained
in the previous section. However, for a large number of atoms and
especially in three dimensions and in the absence of particular
symmetries (e.g. spherical symmetry) this can be a very heavy
numerical task. To overcome this difficulty, in order to develop an
analytical approach, we can exploit the fact that for large $N$ in
the initial state (\ref{eq:initial}) the distributions of the number
of atoms $N_a$ and $N_b$ are very peaked around their average values
with a typical width of order $\sqrt{N}$. Moreover, assuming that
possible fluctuations in the total number of particles are described
by a distribution $P(N)$ having a width much smaller than the
average of the total number of particles $\bar{N}$, we can limit to
$N_a$ and $N_b$ close to $\bar{N}_a=|C_a|^2\bar{N}$ and
$\bar{N}_b=|C_b|^2\bar{N}$. We then split the condensate wave
function into modulus and phase
\begin{equation}
\phi_\varepsilon=|\phi_\varepsilon|\exp(i\theta_\varepsilon) \;
\hspace{0.5cm} \varepsilon=a,b \,,
\end{equation}
and we assume that the variation of the modulus over the
distribution of $N_\varepsilon$ can be neglected while we
approximate the variation of the phase by a linear expansion around
$\bar{N}_\varepsilon$ \cite{EPJD}. The approximate condensate wave
functions read
\begin{equation}\label{modulus}
\phi_\varepsilon(N_a,N_b)\simeq \bar{\phi}_\varepsilon \exp \left
[i\sum_{\varepsilon'=a,b}(N_{\varepsilon'}
-\bar{N}_{\varepsilon'})(\partial_{N_{\varepsilon'}}
\theta_\varepsilon)_{\bar{N}_a,\bar{N}_b}\right ]
\end{equation}
where $\bar{\phi}_\varepsilon\equiv \phi_\varepsilon(N_a=\bar{N}_a,
N_b=\bar{N}_b)$.

The modulus phase approximation takes into account, in an approximate way,
the dependence of the condensate wave functions on the number of
particles. It is precisely this effect that is responsible of
entanglement between spatial dynamics and spin dynamics.

As explained in Appendix \ref{app:modphase}, all
the relevant averages needed to calculate spin squeezing can then be
expressed in terms of $\bar{\phi}_\varepsilon$ and of three time and
position dependent quantities:
\begin{eqnarray}
&&\chi_d(\vec{r})=\frac{1}{2}\left[(\partial_{N_a}-\partial_{N_b})
(\theta_a-\theta_b)\right]_{\bar{N}_a,\bar{N}_b}, \label{eq:chi_d}\\
&&\chi_s(\vec{r})=\frac{1}{2}\left[(\partial_{N_a}+\partial_{N_b})
(\theta_a-\theta_b)\right]_{\bar{N}_a,\bar{N}_b}, \label{eq:chi_s} \\
&&\chi_0(\vec{r})=\frac{1}{2}\left[(\partial_{N_a}-\partial_{N_b})
(\theta_a+\theta_b)\right]_{\bar{N}_a,\bar{N}_b} \label{eq:chi_0} \,.
\end{eqnarray}
In some cases (see Sect. \ref{sub:breath}) these quantities can be
explicitly calculated analytically. To calculate the squeezing in
the general case, it is sufficient to evolve a few coupled
Gross-Pitaevskii equations (\ref{eq:GPE}) for different values of
$N_a$, $N_b$, to calculate numerically the derivatives of the phases
appearing in (\ref{eq:chi_d})-(\ref{eq:chi_0}). Although we do not
expect a perfect quantitative agreement with the full numerical
model for all values of parameters, we will see that the analytical
model catches the main features and allows us to interpret simply
the results.

In the particular case of stationary wave functions of the
condensates, the parameters $\chi_d$, $\chi_s$ and $\chi_0$ become
space-independent:
\begin{eqnarray}
\chi_d^{\text{st}}
&=&- \frac{\left[(\partial_{N_a}-\partial_{N_b})
({\mu}_a-{\mu}_b)\right]_{\bar{N}_a,\bar{N}_b}}{2\hbar} \,
t \label{eq:stat_chi_d}\\
\chi_s^{\text{st}}
&=&- \frac{\left[(\partial_{N_a}+\partial_{N_b})
({\mu}_a-{\mu}_b)\right]_{\bar{N}_a,\bar{N}_b}}{2\hbar} \,
t \label{eq:stat_chi_s} \\
\chi_0^{\text{st}}&=&\chi_s^{\text{st}} \,.
\end{eqnarray}
In this case we recover a simple two-mode model. Equations
(\ref{eq:stat_chi_d})-(\ref{eq:stat_chi_s}) will be used in section
\ref{sec:twomodes}. In that contest we will rename
$\chi_d^{\text{st}}/t=-\chi$ and
$\chi_s^{\text{st}}/t=-\tilde{\chi}$ to shorten the notations.

\begin{figure}
\centering
\includegraphics[scale=0.4]{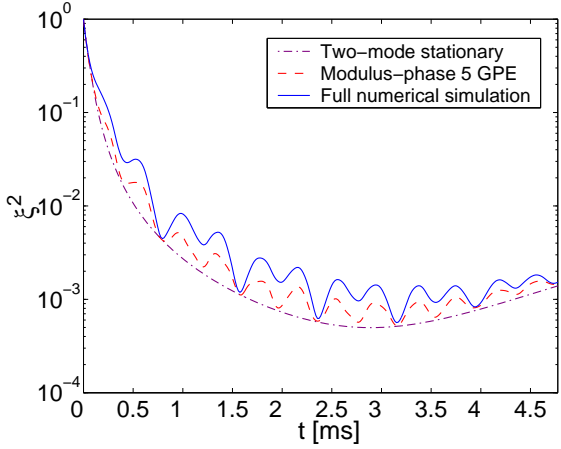}
\caption{Spin squeezing as a function of time. Comparison of the
modulus-phase model (red dashed line) with a full numerical
calculation with 3000 Fock states (blue solid line) and with a
stationary two-mode model (violet dash-dotted line). Spatial
dynamics is strongly excited after the $\pi/2$-pulse populating a
second internal state. $\omega=2\pi \times 2$ kHz,
$\bar{N}_a=\bar{N}_b=5\times10^4$, $m$=87 a.m.u., $a_{aa}=100.44 \,
r_B$, $a_{bb}=95.47 \, r_B$, $a_{ab}=88.28 \, r_B$. No particle
losses. $r_B$ is the Bohr radius.} \label{fig:Rbsteep}
\end{figure}

To test our modulus-phase dynamical model, in Fig.
\ref{fig:Rbsteep}, we consider a situation in which the external
dynamics is significantly excited after the $\pi/2$ pulse which
populates the state $b$. Parameters correspond to a bimodal Rb
condensate in $|F=1,m_F=1\rangle$ and $|F=2,m_F=-1\rangle$ with
$\bar{N}_a= \bar{N}_b=5 \times10^4$ and where a Feshbach resonance
is used to reduce $a_{ab}$ by about 10\% with respect to its bare
value \cite{Sengstock,Widera}. The considered harmonic trap is very
steep $\omega=2\pi \times 2$ kHz. In the figure we compare our
modulus-phase approach (dashed line) with the full numerical
solution (solid line) and with a stationary calculation using
(\ref{eq:stat_chi_d})-(\ref{eq:stat_chi_s}) (dash-dotted line)
which is equivalent to a two-mode model. The
oscillation of the squeezing parameter in the two dynamical
calculations (dashed line and solid line) are due to the fact that
the sudden change in the mean-field causes oscillations in the wave
functions whose amplitude and the frequency are different for each
Fock state. From the figure, we find that our modulus-phase approach
obtained integrating 5 Gross-Pitaevskii equations (dashed line)
reproduces the main characteristics of the full numerical simulation
using 3000 Fock states (solid line). The stationary two mode model
on the other hand is not a good approximation in this case. Only for
some particular times the three curves almost touch. At these times
the wave functions of all the Fock states almost overlap and, as we
will show in our analytical treatment, spatial dynamics and spin
dynamics disentangle.

\begin{figure}
\centering
\includegraphics[scale=0.4]{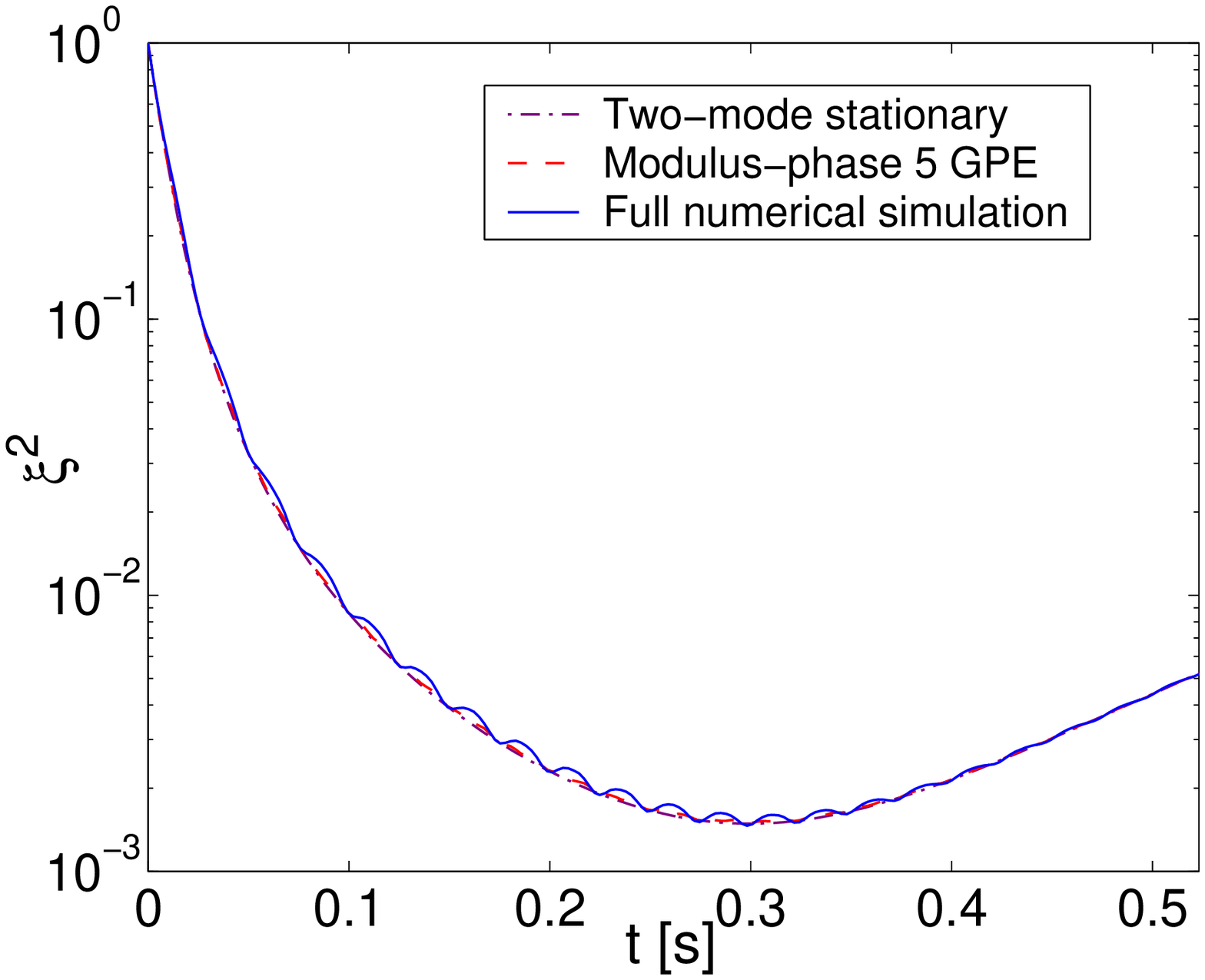} \\
\includegraphics[scale=0.4]{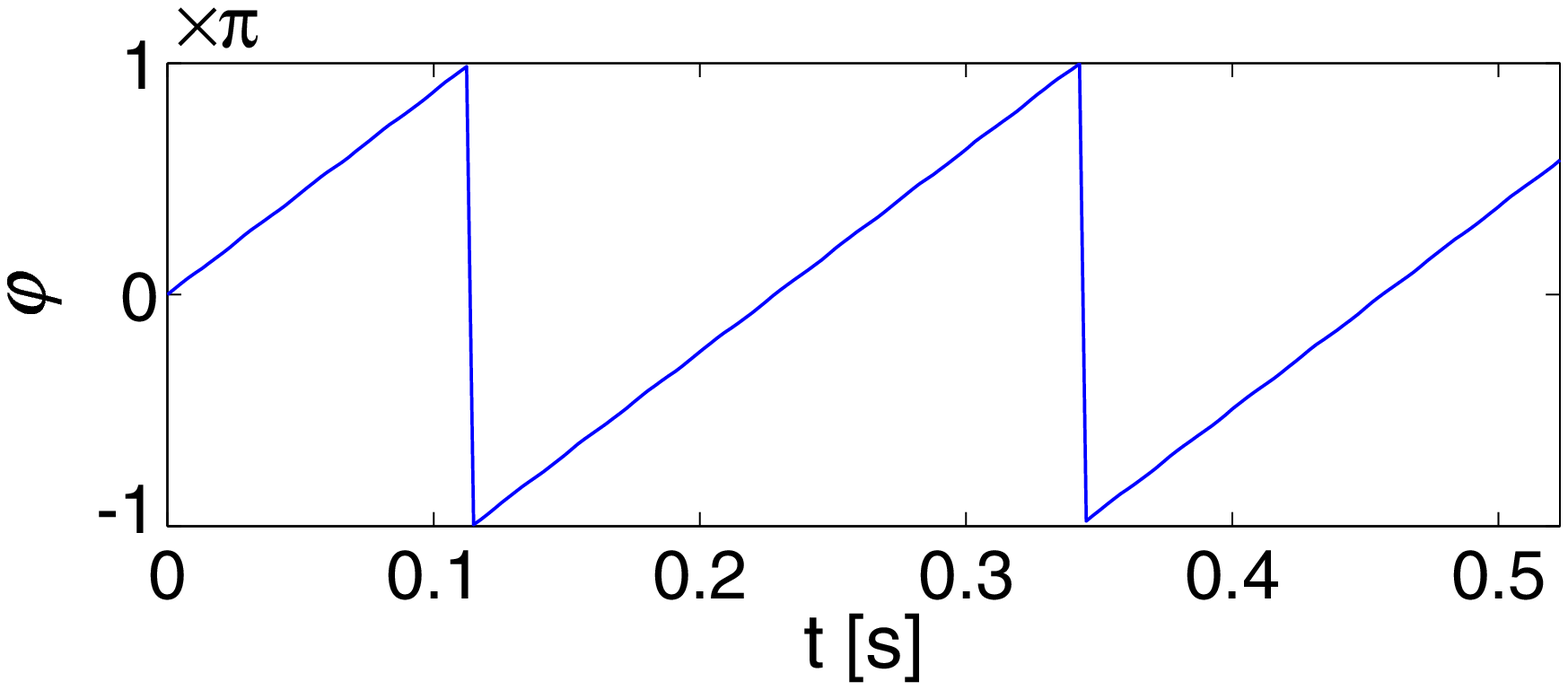}
\caption{(Top) spin squeezing as a function of time in a case in
which the spatial dynamics is weakly excited. Blue solid line: full
numerical calculation with 1000 Fock states. Red dashed line:
modulus-phase model. Violet dash-dotted line: stationary two-mode
model. (Bottom) angle giving the direction of the mean spin
projection in the equatorial plane of the Bloch sphere. Parameters:
$\omega=2\pi \times 42.6$ Hz, $\bar{N}_a=\bar{N}_b=1\times10^4$,
$m$=87 a.m.u., $a_{aa}=100.44 \, r_B$, $a_{bb}=95.47 \, r_B$,
$a_{ab}=88.28 \, r_B$. No particle losses. $r_B$ is the Bohr
radius.} \label{fig:Rbopt}
\end{figure}

In Fig.\ref{fig:Rbopt}  we move to a shallow trap and less
atoms.
We note that in this case both the modulus-phase curve and the
numerical simulation are very close to the stationary two-mode
model which is then a good approximation at all times.

\subsection{Squeezing in the breathe-together solution}
\label{sub:breath}

In this section we restrict to a spherically symmetric harmonic
potential $U^{\mbox{\small ext}}=m\omega^2 r^2/2$ identical for the
two internal sates. For values of the
inter particle scattering lengths such that
\begin{equation}
a_{ab} < a_{aa} , a_{bb}
\end{equation}
and for a particular choice of the mixing angle such that the mean
field seen by the two condensates with $\bar{N}_a$ and $\bar{N}_b$
particles is the same:
\begin{equation}
\bar{N}_a g_{aa} + \bar{N}_b g_{ab} = \bar{N}_b g_{bb} + \bar{N}_a
g_{ab}\equiv \bar{N} g \,, \label{eq:cond_breath}
\end{equation}
the wave functions $\bar{\phi}_a$ and $\bar{\phi}_b$ solve the same
Gross-Pitaevskii equation. In the Thomas-Fermi limit, the wave
functions $\bar{\phi}_a$ and $\bar{\phi}_b$ share the same scaling
solution $\bar{\phi}$ \cite{Scaling1,Scaling2} and ``breathe-together''
\cite{EPJD}.
\begin{equation}
\bar{\phi}_a=\bar{\phi}_b=\bar{\phi}(\vec{r},t)\equiv
\frac{e^{-i\eta(t)}}{{\cal L}^{3/2}(t)} e^{i m r^2 \dot{\cal
L}(t)/2\hbar {\cal L}(t)} \bar{\phi}_0(\vec{r}/{\cal L}(t)\,)
\label{eq:breath}
\end{equation}
with
\begin{eqnarray}
\dot{\eta}&=&\frac{g}{g_{aa}}\frac{\bar{\mu}}{{\cal L}^3\hbar} \\
\frac{d^2{\cal L}}{dt^2}&=&\frac{g}{g_{aa}} \frac{\omega^2}{{\cal
L}^4} -\omega^2 {\cal L} \:;
 \label{eq:lambda}\\
\bar{\phi}_0(\vec{r}\,)&=&\left( \frac{15}{8\pi R_0^3}\right)^{1/2}
\left[ 1- \frac{r^2}{R_0^2}\right]^{1/2}
\end{eqnarray}
$\bar{\mu}$ is the chemical potential of the stationary condensate
{\it before} the $\pi/2$ pulse, when all the $N$ atoms are in state
$a$, and $R_0=\sqrt{2\bar{\mu}/m\omega^2}$ is the corresponding
Thomas-Fermi radius. The initial conditions for (\ref{eq:lambda})
are ${\cal L}(0)=1$ and $\dot{\cal L}(0)=0$.

\begin{figure}[htb]
\centering
\includegraphics[scale=0.4]{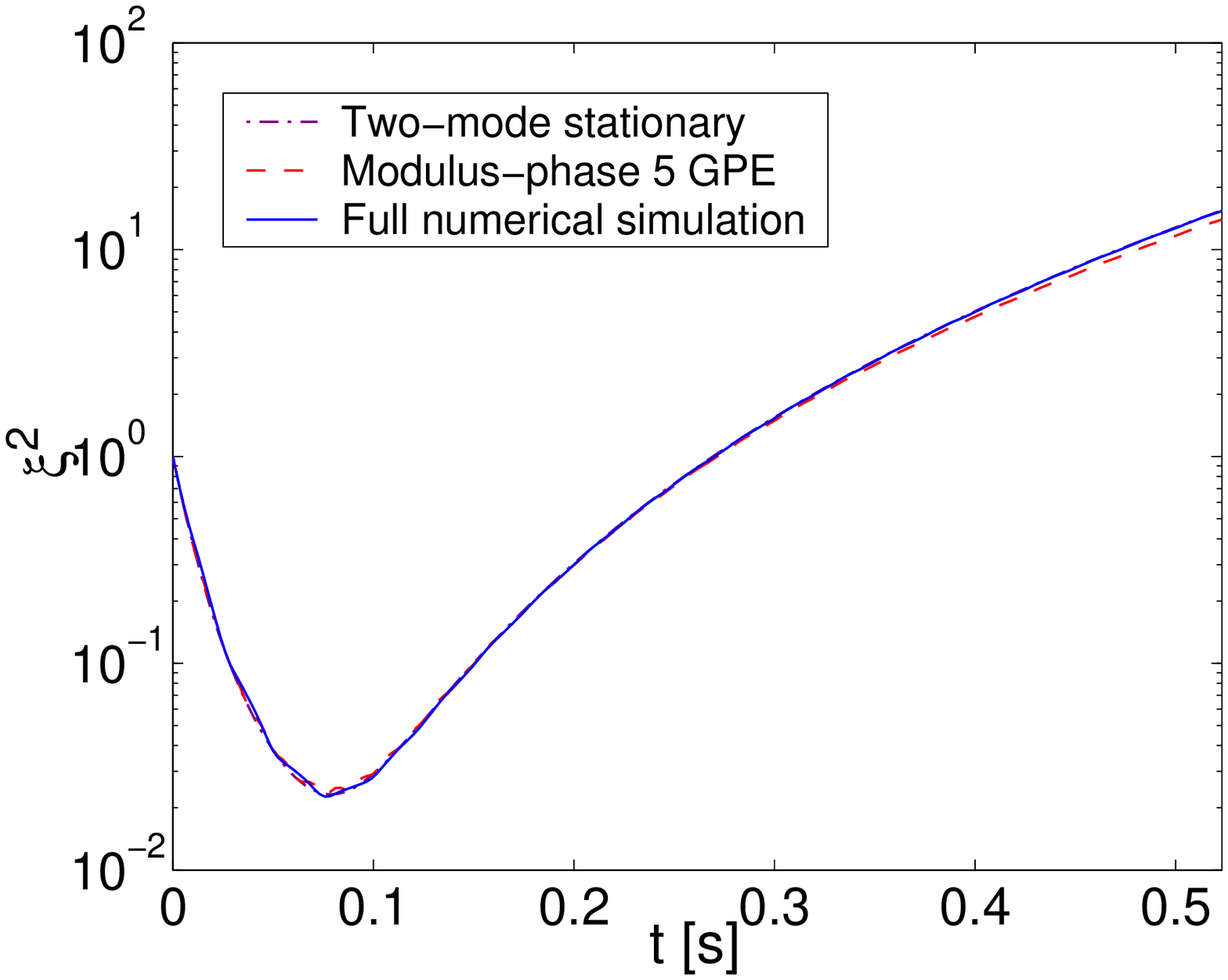}\\
\includegraphics[scale=0.4]{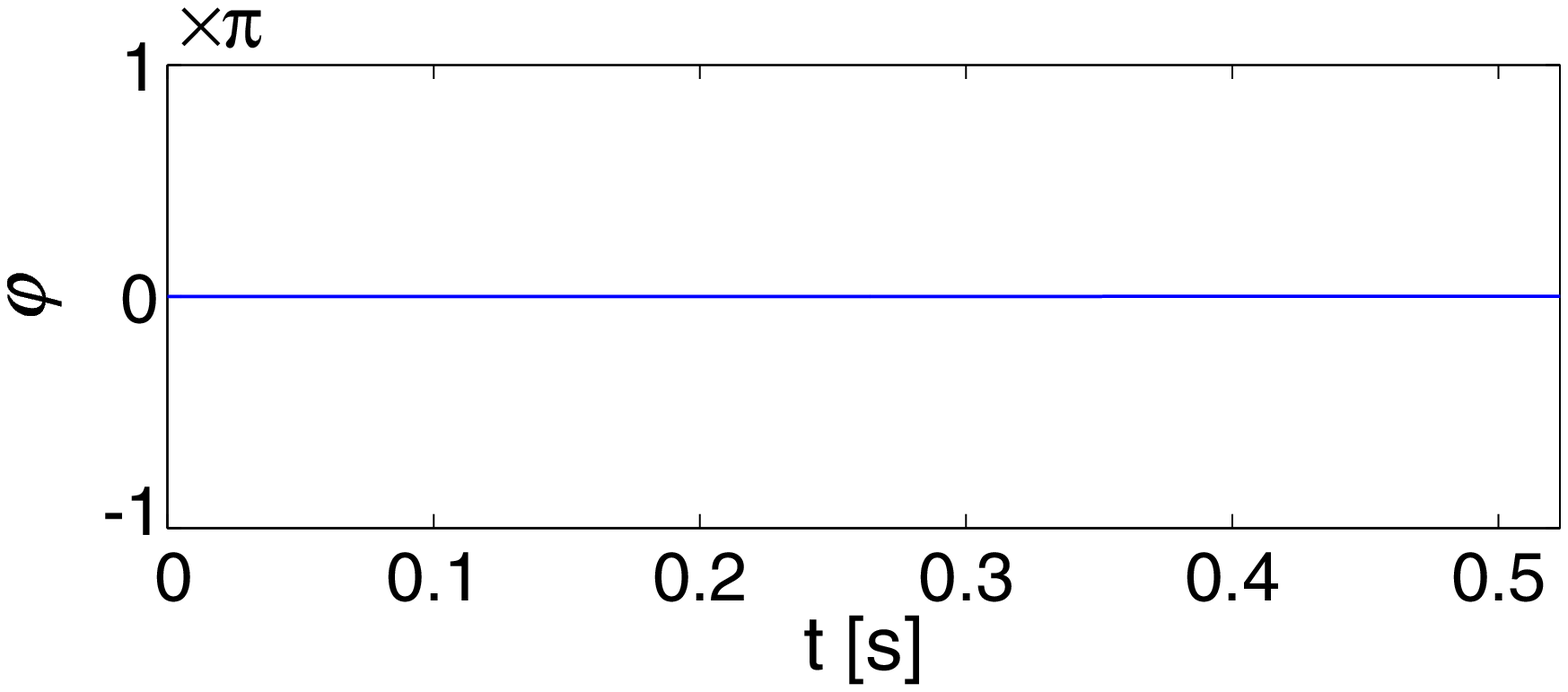}
\caption{(Top) spin squeezing in breathe-together conditions as a
function of time. Blue solid line: full numerical calculation. Red
dashed line: modulus-phase model. Violet dash-dotted line:
stationary two-mode model. (Bottom) angle giving the direction of
the mean spin projection on the equatorial plane of the Bloch
sphere. Parameters: $\omega=2\pi \times 42.6$ Hz, $\bar{N}_a=7432$,
$\bar{N}_b=12568$, $m$=87 a.m.u., $a_{aa}=100.44 \, r_B$,
$a_{bb}=95.47 \, r_B$, $a_{ab}=88.28 \, r_B$. No particle losses.
$r_B$ is the Bohr radius.} \label{fig:Rboptbre}
\end{figure}

Note that the scaling solution identical for the two modes $a$ and
$b$ is valid only for $N_a=\bar{N}_a$, $N_b=\bar{N}_b$ and does not
apply to all the wave functions $\phi_a(N_a,N_b)$ and
$\phi_b(N_a,N_b)$ in the expansion equation (\ref{eq:initial}).
Nevertheless, an advantage of choosing the mixing angle in order to
satisfy the breathe-together condition equation
(\ref{eq:cond_breath}), is that the mean spin has no drift velocity.
In Fig.\ref{fig:Rboptbre} we calculate the spin squeezing (Top) and
the angle $\varphi$ giving the direction of the mean spin projection
on the equatorial plane of the Bloch sphere (Bottom), for the same
parameters as in Fig.\ref{fig:Rbopt} except for the mixing angle
that we now choose satisfying equation (\ref{eq:cond_breath}) while
in Fig.\ref{fig:Rbopt} we had $\bar{N}_a=\bar{N}_b$. Note that
$\varphi$ practically does not evolve. The maximum amount of
squeezing is lower in the breathe-together configuration than in the
even-mixing case (see also \cite{Dutton}). However, as we will see
in the next section, this conclusion does not hold when particle
losses are taken into account.

By linearization of $\phi_a(N_a,N_b)$ and $\phi_b(N_a,N_b)$ around
the breathe-together solution $\bar{\phi}$ and using classical
hydrodynamics, it is even possible to calculate {\it analytically}
the parameters $\chi_d$ and $\chi_s$ relevant for the squeezing
dynamics \cite{EPJD}. One obtains:
\begin{eqnarray}
\chi_d(\vec{r},t)&=& -\frac{1}{2\hbar} \left( \frac{2}{5}
\frac{\bar{\mu}}{N}\right) \frac{g_{aa}+g_{bb}-2g_{ab}}{g_{aa}}\nonumber \\
&& \hspace{-13mm} \times\left\{ \int_0^t \frac{dt'}{{\cal L}^3(t')}
+ \frac{5}{2} \frac{ \mbox{Im} {\cal B}(t)}{\Omega_5} \left[ \left(
\frac{r}{{\cal L}(t)R_0}\right)^2
-\frac{3}{5} \right] \right\} \label{eq:sqBreath}\\
\chi_s(\vec{r},t)&=&\left( |C_b|^2-|C_a|^2\right)
\chi_d=\chi_0(\vec{r},t) \,.
\end{eqnarray}
with
\begin{equation}
\Omega_5=\left( \frac{\bar{N}_a \bar{N}_b}{N^2}
\frac{g_{aa}+g_{bb}-2g_{ab}}{g_{aa}} \right)^{1/2} 5^{1/2} \omega
\end{equation}
and where $\mbox{Im} {\cal B}(t)$ is solution of the differential
equations
\begin{eqnarray}
i \dot{{\cal{A}}} &=& \frac{\Omega_5}{{\cal L}^2(t)} {\cal{B}} \\
i \dot{{\cal{B}}} &=& \frac{\Omega_5}{{\cal L}^3(t)} {\cal{A}}
\end{eqnarray}
to be solved together with equation (\ref{eq:lambda}), with initial
conditions ${\cal A}(0)={\cal B}(0)=1$. In practice, when we expand
the condensate wave functions around the breathe-together solution
equation (\ref{eq:breath}) as in \cite{EPJD}, we encounter the
hydrodynamics operator $S$ \cite{Stringari}
\begin{equation}
S[\alpha] \equiv -\frac{Ng_{aa}}{M} \mbox{div} [\bar{\phi}_0^2 \:
{\mbox{\bf grad}} \,\alpha] \,.
\end{equation}
The deviation of the relative phase and the relative density from
the breathe-together solution expand over two eigenmodes of $S$: A
zero-energy mode which grows linearly in time and gives the dominant
features of phase dynamics and squeezing (integral term in the curly
brackets in Eq.(\ref{eq:sqBreath})), and a breathing mode of
frequency $\Omega_5$ which is responsible for the oscillations of
the squeezing parameter. The fact that in breathe-together
conditions and within the modulus-phase approximation
$\chi_s=\chi_0$ is shown in Appendix \ref{app:chis_chi0}.

\begin{figure}
\centering
\includegraphics[scale=0.4]{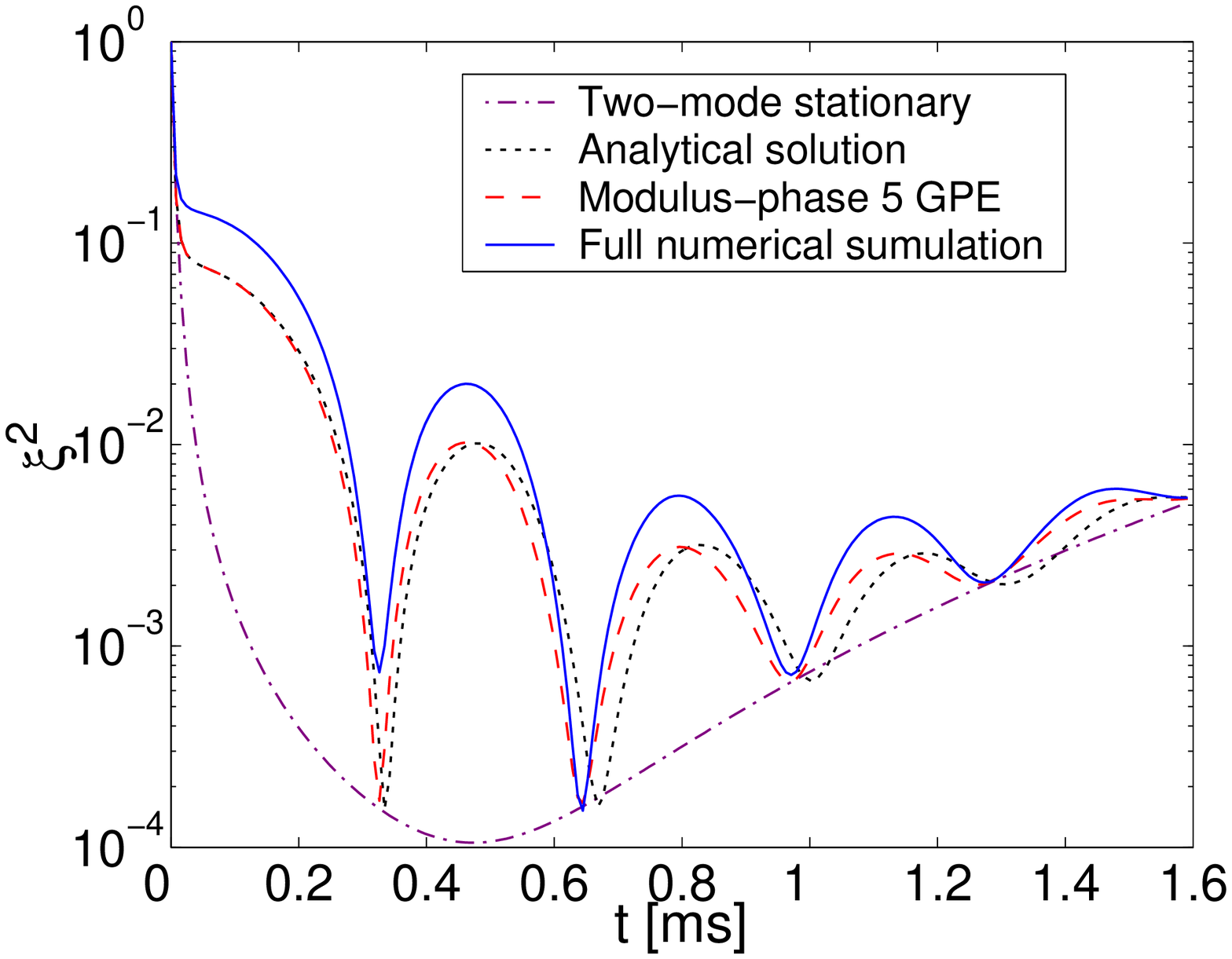}\\
\includegraphics[scale=0.4]{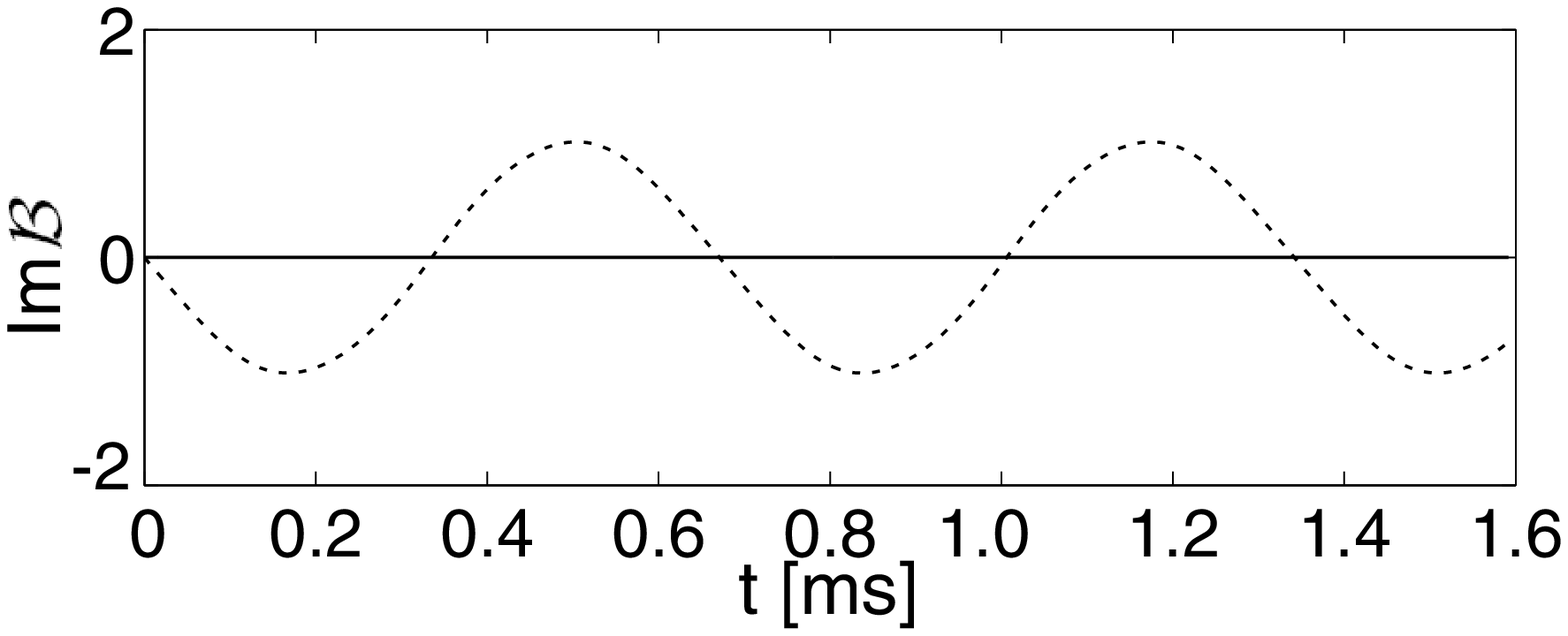}
\caption{(Top) test of the analytical formula equation
(\ref{eq:sqBreath}) in the deep Thomas-Fermi regime. Spin squeezing
as a function of time. Blue solid line: full numerical calculation.
Red dashed line: modulus-phase model. Black dotted line: analytical
curve using equation (\ref{eq:sqBreath}). Violet dash-dotted line:
stationary two-mode model using
(\ref{eq:stat_chi_d})-(\ref{eq:stat_chi_s}). (Bottom) function
Im${\cal B}(t)$. Spatial and spin dynamics disentangle when Im${\cal
B}(t)=0$. Parameters: $\bar{N}_a=\bar{N}_b=5\times 10^5$,
$\omega=2\pi \times 2$ kHz, $m$=87 a.m.u., $a_{aa}=a_{bb}=0.3
\,a_{ho}$, $a_{ab}=0.24 \,a_{ho}$. $a_{ho}$ is the harmonic
oscillator length: $a_{ho}=\sqrt{\hbar/M\omega}$. No particle
losses.} \label{fig:analy}
\end{figure}

We give an example corresponding to strongly oscillating wave
functions in Fig.\ref{fig:analy} where we compare the spin squeezing
from the analytical theory with a numerical simulation. In the
analytical formula, the entanglement between spatial degrees of
freedom and spin dynamics is apparent as $\chi_d$ equation
(\ref{eq:sqBreath}) is position dependent. The points in which the
dynamical curve (dotted line) touches the stationary two mode curve
(dash-dotted line) correspond to Im${\cal B}=0$ (see the bottom
curve) where space and spin dynamics are disentangled. We note
however that the validity conditions of classical hydrodynamics are
more stringent for a mixture of condensates with rather close
scattering lengths than for a single condensate \cite{EPJD}. We
checked numerically that in order for equation (\ref{eq:sqBreath})
to correctly predict the frequency of the oscillations in the
squeezing parameter, we have to enter deeply in the Thomas-Fermi
regime.

\subsection{``Extracted'' spin squeezing}
\label{sub:extract}

As we pointed out, the definitions equations
(\ref{eq:Sx})-(\ref{eq:Sz}) explicitly include the spatial overlap
between the two modes. Here we give an alternative definition that
can be used always, whether or not the modes overlap. To this aim,
we introduce the time-dependent operators
\begin{eqnarray}
\tilde{a}&=&\int d^3r \: \bar{\phi}^\ast_a(\vec{r},t)
\hat{\psi}_a(\vec{r}),\;  \label{eq:extraa}\\
\tilde{b}&=&\int d^3r \:  \bar{\phi}^\ast_b(\vec{r},t)
\hat{\psi}_b(\vec{r}),\; \label{eq:extrbb}
\end{eqnarray}
where $\bar{\phi}_\varepsilon(\vec{r},t)$ is the solution of
Gross-Pitaevskii equation (\ref{eq:GPE}) for mode $\varepsilon$ with $\bar{N}_a$,
$\bar{N}_b$ particles. We then introduce the spin operators:
\begin{eqnarray}
\tilde{S}_x=&\dfrac{1}{2}(\tilde{b}^\dag\tilde{a}+\tilde{a}^\dag
\tilde{b}),\label{spin_internal_x}\\
\tilde{S}_y=&\dfrac{i}{2}(\tilde{b}^\dag\tilde{a}-\tilde{a}^\dag
\tilde{b}),\label{spin_internal_y}\\
\tilde{S}_z=&\dfrac{1}{2}(\tilde{a}^\dag\tilde{a}-\tilde{b}^\dag
\tilde{b}).\label{spin_internal_z}
\end{eqnarray}
In the new definition of spin squeezing calculated by the spin
operators defined in equations
(\ref{spin_internal_x})-(\ref{spin_internal_z}), which we call the
``extracted'' spin squeezing, we still take into account
entanglement between external motion and spin dynamics, but we give
up the information about the overlap between the two modes. In
Appendix \ref{app:extracted}, we give the quantum averages useful to
calculate the extracted spin squeezing within the modulus-phase
approach described in Section \ref{sub:modulusphase}. We will use
this extracted spin squeezing in Section \ref{sec:Galette}.

Comparing the expressions given in Appendix \ref{app:extracted} with
those of Appendix \ref{app:twomodes} (in the absence of losses), one
realizes that in the stationary case, where $\chi_d$, $\chi_s$ and
$\chi_0$ are space independent, the extracted spin squeezing
dynamical model reduces to a two-mode model that we study in detail
in the next section.

\section{Two-mode model with Particle losses}
\label{sec:twomodes}

In this section we generalize our results of \cite{PRLlosses} to
possibly overlapping and non-symmetric condensates. In subsection
\ref{sub:lossasym} we address the general case, while in subsection
\ref{sub:symm} we restrict to symmetric condensates and perform
analytically an optimization of the squeezing with respect to the
trap frequency and number of atoms. In the whole section, as in
\cite{PRLlosses}, we will limit to a two-mode stationary model and
we do not address dynamical issues.

\subsection{Spin squeezing in presence of losses}
\label{sub:lossasym}

We consider a two-component Bose-Einstein condensate initially
prepared in a phase state, that is with well defined relative phase
between the two components,
\begin{equation}
|\Psi(0)\rangle=|\varphi \rangle \equiv \frac{\left(
|C_a|e^{-i\varphi/2} a^\dag+ |C_b| e^{i\varphi/2} b^\dag
\right)^N}{\sqrt{N!}}|0\rangle \,. \label{eq:initial_twomodes}
\end{equation}
When expanded over Fock states, the state
(\ref{eq:initial_twomodes}) shows binomial coefficients which, for
large $N$, are peaked around the average number of particles in $a$
and $b$, $\bar{N}_a$ and $\bar{N}_b$. In the same spirit as the
``modulus-phase'' approximation of subsection
\ref{sub:modulusphase}, we can use this fact to expand the
Hamiltonian of the system to the second order around $\bar{N}_a$ and
$\bar{N}_b$
\begin{eqnarray}
H_0&\simeq & E(\bar{N}_a,\bar{N}_b)+\sum_{\varepsilon=a,b}
\mu_\varepsilon(\hat{N}_\varepsilon-\bar{N}_\varepsilon)
+\frac{1}{2}\partial_{N_\varepsilon}
\mu_\varepsilon(\hat{N}_\varepsilon-\bar{N}_\varepsilon)^2 \nonumber \\
&+&\frac{1}{2}\left(\partial_{N_b}\mu_a+\partial_{N_a}\mu_b\right)
(\hat{N}_a-\bar{N}_a) (\hat{N}_b-\bar{N}_b)
\end{eqnarray}
where the chemical potentials $\mu_\varepsilon$ and all the
derivatives of $\mu_\varepsilon$ should be evaluated in $\bar{N}_a$
and $\bar{N}_b$. We can write
\begin{equation}
H_0=f_{\hat{N}}+ \hbar v_{\hat{N}} (\hat{N}_a-\hat{N}_b)
+\frac{\hbar\chi}{4}(\hat{N}_a-\hat{N}_b)^2
\label{eq:hamitonian}
\end{equation}
with
\begin{eqnarray}
v_{\hat{N}} &=& \frac{1}{2\hbar} \left[ (\mu_a-\mu_b) - \hbar \chi (\bar{N}_a-\bar{N}_b)
+ \hbar \tilde{\chi} (\hat{N}-\bar{N})\right] \label{eq:vdrift} \\
\chi&=&\frac{1}{2\hbar}\left(\partial_{N_a}\mu_a+\partial_{N_b}\mu_b
-\partial_{N_b}\mu_a-\partial_{N_a}\mu_b\right)_{\bar{N}_a,\bar{N}_b}
\,  \label{eq:chi2M} \\
\tilde{\chi} &=& \frac{1}{2\hbar}(\partial_{N_a} \mu_a
- \partial_{N_b} \mu_b)_{\bar{N}_a,\bar{N}_b} \,. \label{eq:tildechi2M}
\end{eqnarray}
The function $f$ of the total number of particles,
$\hat{N}=\hat{N}_a + \hat{N}_b$, commutes with the density operator
of the system and can be omitted. The second term in equation
(\ref{eq:hamitonian}) proportional to $S_z$ describes a rotation of
the average spin vector around the $z$ axis with velocity
$v_{\hat{N}}$. The third term proportional to $S_z^2$ provides the
nonlinearity responsible for spin squeezing. It also provides a
second contribution to the drift of the relative phase between the
two condensates in the case $\bar{N}_a\neq\bar{N}_b$.

In presence of losses, the evolution is ruled by a master equation
for the density operator $\rho$ of the system. In the interaction
picture with respect to $H_0$, with one, two, and three-body losses,
we have:
\begin{eqnarray}
\frac{d\tilde{\rho}}{dt}&=&\sum_{m=1}^{3} \sum_{\varepsilon=a,b}
\gamma_\varepsilon^{(m)}\left[{c}^m_\varepsilon\tilde{\rho}{c}^{\dag
m}_\varepsilon-\frac{1}{2}\{{c}^{\dag m}_\varepsilon{c}^m_
\varepsilon,
\tilde{\rho}\} \right] \nonumber \\
&+&\gamma_{ab}\left[c_a c_b \tilde{\rho} c_a^\dag
c_b^\dag-\frac{1}{2}\{c_a^\dag c_b^\dag c_a c_b,
\tilde{\rho}\}\right]
\end{eqnarray}
where $\tilde{\rho}=e^{iH_0t/\hbar} \rho e^{-iH_0t/\hbar}$,
${c}_a=e^{iH_0t/\hbar} a e^{-iH_0t/\hbar}$, and similarly for $b$,
\begin{eqnarray}
\gamma_\varepsilon^{(m)}&=&\frac{K_m^{(\varepsilon)}}{m}\int
d^3r|\bar{\phi}_\varepsilon(r)|^{2m}\, ,\\
 \gamma_{ab}&=&\frac{K_{ab}}{2}\int
d^3r|\bar{\phi}_a(r)|^{2}|\bar{\phi}_b(r)|^{2} \,.
\end{eqnarray}
$K_m^{(\varepsilon)}$ is the $m$-body rate constant ($m=1,2,3$) and
$\bar{\phi}_\varepsilon(r)$ is the condensate wave function for the
$\varepsilon$ component with $N_a=\bar{N}_a$ and $N_b=\bar{N}_b$
particles. $K_{ab}$ is the rate constant for a two-body loss event
in which two particles coming from different components are lost at
once.

In the Monte Carlo wave function approach \cite{MCD} we define an
effective Hamiltonian $H_{\text{eff}}$ and the jump operators
$J_\varepsilon^{(m)}$($J_{ab}^{(2)}$)
\begin{eqnarray}
& & H_{\text{eff}}=-\frac{i\hbar}{2} \sum_{m=1}^{3}
\sum_{\varepsilon=a,b} \gamma_{\varepsilon}^{(m)} c_
\varepsilon^{\dag m}c_\varepsilon^m -\frac{i\hbar}{2}
\gamma_{ab}c_a^\dag c_b^\dag c_a c_b\,; \label{H_eff} \\
& &
J_\varepsilon^{(m)}=\sqrt{\gamma_{\varepsilon}^{(m)}}{c}^m_\varepsilon\,,
\; J_{ab}^{(2)}=\sqrt{\gamma_{ab}}c_a c_b
\end{eqnarray}
We assume that a small fraction of particles will be lost during the
evolution so that we can consider $\chi$,
$\gamma_{\varepsilon}^{(m)}$ and $\gamma_{ab}$ as constant
parameters of the model. The state evolution in a single quantum
trajectory is a sequence of random quantum jumps at times $t_j$ and
non-unitary Hamiltonian evolutions of duration $\tau_j$:
\begin{eqnarray}
|\Psi(t)\rangle&=&e^{-iH_{\text{eff}} (t-t_k)/\hbar}J^{(m_k)}_{
\varepsilon_k}(t_k)e^{-iH_{\text{eff}} \tau_k/\hbar}J^{(m_{k-1})}_{
\varepsilon_{k-1}}(t_{k-1})\nonumber \\
&&\ldots J^{(m_1)}_{\varepsilon_1}(t_1)e^{-iH_{\text{eff}}
\tau_1/\hbar}|\Psi(0)\rangle\,, \label{eq:evol}
\end{eqnarray}
where now $\varepsilon_j=a,b$ or $ab$. Application of a jump
$J^{(m_j)}_{\varepsilon_j} (t_j)$ to the $N$-particle phase state at
$t_j$ yields
\begin{eqnarray}
&&J^{(m_j)}_{\varepsilon_j}(t_j)|\phi\rangle_N \propto
|\phi +  \Delta_j t_j \rangle_{N-m_j} \,, \label{eq:jump_on_phasestate}\\
&&\Delta_j = 2 \tilde{\chi} \delta_{\varepsilon_j,ab} +
(\tilde{\chi} + \chi) m_j  \delta_{\varepsilon_j,a}
+ (\tilde{\chi} - \chi) m_j \delta_{\varepsilon_j,b} \,.
\end{eqnarray}
After a quantum jump, the phase state is changed into a new phase
state, with $m$ particle less and with the relative phase between
the two modes showing a random shift $\Delta_j t_j$ with respect to
the phase before the jump. Note that in the symmetrical case
$\tilde{\chi}=0$ and no random phase shift occurs in the case of a
jump of $ab$. Indeed we will find that at short times in the
symmetrical case theses kind of crossed $ab$ losses are harmless to
the the squeezing.

In presence of one-body losses only, also the effective Hamiltonian
changes a phase state into another phase state and we can calculate
exactly the evolution of the state vector analytically, as we did in
\cite{PRLlosses} for symmetrical condensates. When two and
three-body losses enter into play, we introduce a constant loss rate
approximation \cite{EPJDloss}
\begin{equation}\label{H_eff_app}
H_{\text{eff}}\simeq - \frac{i\hbar}{2}\sum_{m=1}^{3}
\sum_{\varepsilon=a,b} \gamma_{\varepsilon}^{(m)}
\bar{N}_\varepsilon^{m} -
\frac{i\hbar}{2}\gamma_{ab}\bar{N}_a\bar{N}_b\equiv -
\frac{i\hbar}{2}\lambda
\end{equation}
valid when a small fraction of particles is lost at the time at
which the best squeezing is achieved. In this approximation, the
mean number of particles at time $t$ is
\begin{eqnarray}
& &\langle\hat{N}\rangle=N\left[1-\left(\sum_{\varepsilon=a,b}
\sum_m\Gamma_{\varepsilon}^{(m)} +\Gamma_{ab}\right)t\right]\\
& &\Gamma_{\varepsilon}^{(m)}\equiv \bar{N}_\varepsilon^{m-1}
m\gamma_\varepsilon^{(m)}\;;
\;\Gamma_{ab}=\gamma_{ab}\sqrt{\bar{N}_a \bar{N}_b}
\end{eqnarray}
where for example $\Gamma_{\varepsilon}^{(m)} t$ is the fraction of
lost particles due to $m$-body losses in the $\varepsilon$
condensate. Let us present the evolution of a single quantum
trajectory: Within the constant loss rate approximation, we can move
all the jump operators in (\ref{eq:evol}) to the right. We obtain:
\begin{eqnarray}
|\Psi(t)\rangle &=& e^{-\lambda t/2}  \prod_{j=1}^{k}
J_{\varepsilon_j}^{(m_j)} (t_j) |\Psi(0)\rangle \nonumber \\
&  = & e^{-\lambda t/2} e^{-i T_k} |\alpha_k| |\varphi + \beta_k
\rangle_{N-N(k)} \label{eq:onetraj}
\end{eqnarray}
where
\begin{eqnarray}
|\alpha_k|^2 & = & \prod_{j=1}^k \left\{ \sum_{m'=1,2,3}
\sum_{\varepsilon'=a,b} \delta_{m_j,m'} \delta_{\varepsilon_j,
\varepsilon'} \bar{N}^{m'} |C_{\varepsilon'}|^{2m'}
\gamma_{\varepsilon'}^{(m')}
\right. \nonumber \\
&+& \left. \bar{N}^2 |C_a|^2 |C_b|^2 \gamma_{ab}
\delta_{\varepsilon_j,ab}  \right\} \label{eq:alpha2}\\
\beta_k &=& \sum_{j=1}^k  t_j \left\{ 2 \tilde{\chi}
\delta_{\varepsilon_j,ab} + \sum_{m'=1,2,3}  m' \delta_{m_j,m'}
\right. \nonumber \\
&\times& \left. [(\chi + \tilde{\chi}) \delta_{\varepsilon_j,a}-
(\chi - \tilde{\chi}) \delta_{\varepsilon_j,b}]
\right\} \\
N(k)&=&  \sum_{j=1}^k \sum_{m'=1,2,3} m' \delta_{m_j,m'}
(\delta_{\varepsilon_j,a}+\delta_{\varepsilon_j,b})+2
\delta_{\varepsilon_j,ab}\,,
\end{eqnarray}
and $T_k$ is a phase which cancels out when taking the averages of
the observables \cite{note_alpha}.

The expectation value of any observable $\hat{\mathcal {O}}$ is
obtained by averaging over all possible stochastic realizations,
that is all kinds, times and number of quantum jumps, each
trajectory being weighted by its probability
\begin{equation}
\langle\hat{\mathcal {O}}\rangle=\sum_{k}\int_{0<t_1<t_2<\cdots
t_k<t} \hspace{-1.5cm} dt_1dt_2\cdots dt_k \sum_{\{\varepsilon_j,
m_j\}} \langle \Psi(t)|\hat{\mathcal {O}}|\Psi(t)\rangle\,.
\end{equation}
Note that the single trajectory (\ref{eq:onetraj}) is not
normalized. The prefactor will provide its correct ``weight'' in the
average.

We report in Appendix \ref{app:1BL} and \ref{app:twomodes} the
averages needed to calculate the spin squeezing for one-body losses
only (exact solution) and for one, two and three-body losses
(constant loss rate approximation) respectively. The analytical
results are expressed in terms of the parameters $\chi$ and
$\tilde{\chi}$ defined in equations (\ref{eq:chi2M}) and
(\ref{eq:tildechi2M}) respectively and of the drift velocity
\begin{equation}
v = \frac{1}{2\hbar} \left[ (\mu_a-\mu_b) - \hbar \chi (\bar{N}_a-\bar{N}_b)
+ \hbar \tilde{\chi} (N-\bar{N})\right] \label{eq:v}\,,
\end{equation}
where $N$ is the total initial number of atoms.

\subsection{Symmetrical case: optimization of spin squeezing}
\label{sub:symm}

If we restrict to symmetrical condensates which may or may not
overlap, we can carry out analytically the optimization of squeezing
in presence of losses. In the symmetric case and constant loss rate
approximation it turns out that $\Delta S^2_z=\langle \hat{N}
\rangle /4$. This allows to express $\xi^2$ in a simple way:
\begin{equation}
\xi^2=\frac{\langle a^\dagger a\rangle}{\langle b^\dagger a
\rangle^2}\left( \langle a^\dagger a\rangle + {\tilde{A}} -
\sqrt{{\tilde{A}}^2+{\tilde{B}}^2}\right)\,, \label{eq:def2xi2}
\end{equation}
with
\begin{eqnarray}
\tilde{A} &=& \frac{1}{2}\, \mbox{Re}\left( \langle b^\dagger a^\dagger a b
- b^\dagger b^\dagger a a\rangle \right) \label{eq:A} \\
\tilde{B} &=& 2 \; \mbox{Im} \left( \langle b^\dagger b^\dagger b a \rangle
\right) \,. \label{eq:B}
\end{eqnarray}
An analytical expression for spin squeezing is calculated from
(\ref{eq:def2xi2}) with
\begin{eqnarray}
& &\langle b^\dag a\rangle=\frac{e^{-\lambda t}}{2}\cos^{N-1}(\chi
t) \tilde{N}F_1 \label{eq:ab}\\
& & \tilde{A}=\frac{e^{-\lambda t}}{8} \tilde{N}(\tilde{N}-1)
\left[ F_0 - F_2 \cos^{N-2}(2\chi t)\right] \label{eq:A2}\\
& &\tilde{B}=\frac{e^{-\lambda t}}{2} \cos^{N-2}(\chi t) \sin(\chi t)
\tilde{N}(\tilde{N}-1) F_1\label{eq:B2}
\end{eqnarray}
where the operator $\tilde{N}=(N-\partial_\sigma)$ acts on the
functions
\begin{eqnarray}
F_\beta(\sigma)&=&\exp\left[ \sum_{m=1}^3 2 \gamma^{(m)}t e^{\sigma
m}\frac{\sin (m \beta \chi t)}{m\beta \chi t \cos^m(\beta \chi
t)}  \right. \nonumber \\
&&+\left. \frac{\gamma_{ab}te^{2\sigma}}{\cos^2(\beta \chi t)}\right] \,,
\end{eqnarray}
with $\beta=0,1,2$ and all expressions should be evaluated in
$\sigma=\ln \bar{N}$.

We want now to find simple results for the best squeezing and the
best squeezing time in the large $N$ limit. In the absence of losses
\cite{Ueda} the best squeezing and the best squeezing time in units
of $1/\chi$ scale as $N^{-2/3}$. We then set $N=\varepsilon ^{-3}$
and rescale the time as $\chi t=\tau \varepsilon^2$. We expand
(\ref{eq:def2xi2}) for $\varepsilon \ll 1$ up to order 2 included,
keeping ${\Gamma}^{(m)}/\chi$ constant. The key point is that in
this expansion, for large $N$ and short times, the crossed losses
$ab$ do not contribute. As in \cite{PRLlosses}, introducing the
squeezing $\xi^2_0(t)$ in the no-loss case, we obtain:
\begin{equation}
\xi^2(t)=\xi^2_0(t)\left[1+\frac{1}{3}
\frac{\Gamma_{\text{sq}}t}{\xi^2_0(t)}\right]\,.
\label{eq:textbook}
\end{equation}
with:
\begin{equation}
\Gamma_{\text{sq}}=\sum_m \Gamma_{\text{sq}}^{(m)} \,\, \mbox{and}
\, \, \Gamma_{\text{sq}}^{(m)} = m {\Gamma}^{(m)}\,.
\end{equation}
The result (\ref{eq:textbook}) very simply accesses the impact of
losses on spin squeezing. First it shows that losses cannot be
neglected as soon as the lost fraction of particles is of the order
of $\xi^2_0$. Second it shows that in the limit $N\to \infty$ and
$\xi^2_0(t_{\text{best}}) \to 0$, the squeezing in presence of
losses is of the order of the lost fraction of particles at the best
time: $\xi^2(t_{\text{best}}) \sim
\Gamma_{\text{sq}}t_{\text{best}}/3$. This also sets the limits of
validity of our constant loss rate approximation. For our
approximation to be valid, the lost fraction of particle, hence
squeezing parameter at the best squeezing time, should be small.

From now on, the optimization of the squeezing in the large $N$
limit proceeds much as in the case of spatially separated
condensates \cite{PRLlosses}. The only difference is in the
stationary wave functions in the Thomas-Fermi limit. For overlapping
condensates we consider a stable mixture with
\begin{equation}
a_{ab} < a_{aa}=a_{bb} \,,
\end{equation}
and we introduce the {\it sum} and {\it difference} of the intra and
inter-species $s$-wave scattering lengths:
\begin{eqnarray}
a_s&=&a_{aa}+a_{ab} \label{eq:a_s}\\
a_d&=&a_{aa}-a_{ab}  \label{eq:a_d}\, .
\end{eqnarray}
In the symmetric case considered here we have
\begin{eqnarray}
&&\mu_a=\mu_b=\frac{1}{2}\hbar\bar{\omega}\left[\frac{15}{2}
\frac{N a_s}{a_{ho}}\right]^{2/5}\,, \\
&&\chi=\frac{2^{3/5}3^{2/5}}{5^{3/5}}\left(\frac{\hbar}{M}
\right)^{-1/5}\bar{\omega}^{6/5}N^{-3/5}\frac{a_d}
{a_s^{3/5}}\\
&&\Gamma^{(1)}=K_1\\
&&\Gamma^{(2)}=\frac{15^{2/5}}{2^{7/5}7\pi}\left(\frac{\hbar}
{M}\right)^{-6/5} \bar{\omega}^{6/5}N^{2/5}a_s^{-3/5}K_2\\
&&\Gamma^{(3)}= \frac{5^{4/5}}{2^{19/5}3^{1/5}7 \pi^2}
\left(\frac{\hbar}{M}\right) ^{-12/5} \bar{ \omega} ^{12/5}
N^{4/5} \nonumber \\
&&\hspace{1cm} a_s^{-6/5} K_3 \,,
\end{eqnarray}
where $a_{ho}=\sqrt{\hbar/M\bar{\omega}}$\, is the harmonic
oscillator length, $\bar{\omega}$ is the geometric mean of the trap
frequencies. We recover the case of spatially separated condensates
\cite{PRLlosses} setting $a_{ab}=0$ in
(\ref{eq:a_s})-(\ref{eq:a_d}).

The squeezing parameter for the best squeezing time
$\xi^2(t_{\text{best}},\bar{\omega})$ is minimized for an optimized
trap frequency
\begin{equation}
\bar{\omega}^{\text{opt}}= \frac{2^{19/12}7^{5/12}\pi^{5/6}}
{15^{1/3}} \frac{\hbar}{M}\frac{a_s^{1/2}}{N^{1/3}}
\left(\frac{K_1}{K_3}\right)^{5/12}. \label{eq:omopt}
\end{equation}
Note however that this optimization concerns one- and three-body
losses only. The effect of decoherence due two two-body losses
quantified by the ratio $\Gamma^{(2)}/\chi$ is independent of the
trap frequency.

Once the trap frequency is optimized,
$\xi^2(t_{\text{best}},\bar{\omega}^{\text{opt}})$ is a decreasing
function of $N$. The lower bound for $\xi^2$, reached for $N
\rightarrow \infty$ is then
\begin{equation}
\inf_{t,\bar{\omega},N} \xi^2 = \left[ \frac{5\sqrt{3}M}{28\pi\hbar
}\right]^{2/3}\left[\sqrt{\frac{7}{2}\frac{K_1
K_3}{a_d^2}}+\frac{K_2} {a_d} \right]^{2/3}. \label{eq:lboundxi2}
\end{equation}

A simple outcome of this analytic study is that, for positive
scattering lengths $a_{aa}$, $a_{ab}$, the maximum squeezing is
obtained when $a_{ab}=0$ that is for example for spatially separated
condensates. Another possibility is to use a Feschbach resonance to
decrease the inter-species scattering length $a_{ab}$
\cite{Sengstock,Widera}, knowing that the crossed $a-b$ losses do
not harm the squeezing at short times.

\section{Results for overlapping condensates}
\label{sec:results_over}
In this and the next section we give practical examples of
application of the analysis led in the two previous sections.

\subsection{Feshbach resonance-tuned bimodal Rb BEC}
We consider a bimodal Rb condensate in $|F=1,m_F=1\rangle$ and
$|F=2,m_F=-1\rangle$ states where the scattering length $a_{ab}$ is
lowered by about 10\% with respect to its bare value using a
Feshbach resonance \cite{Sengstock,Widera}.

\begin{figure}
\centering
\includegraphics[scale=0.4]{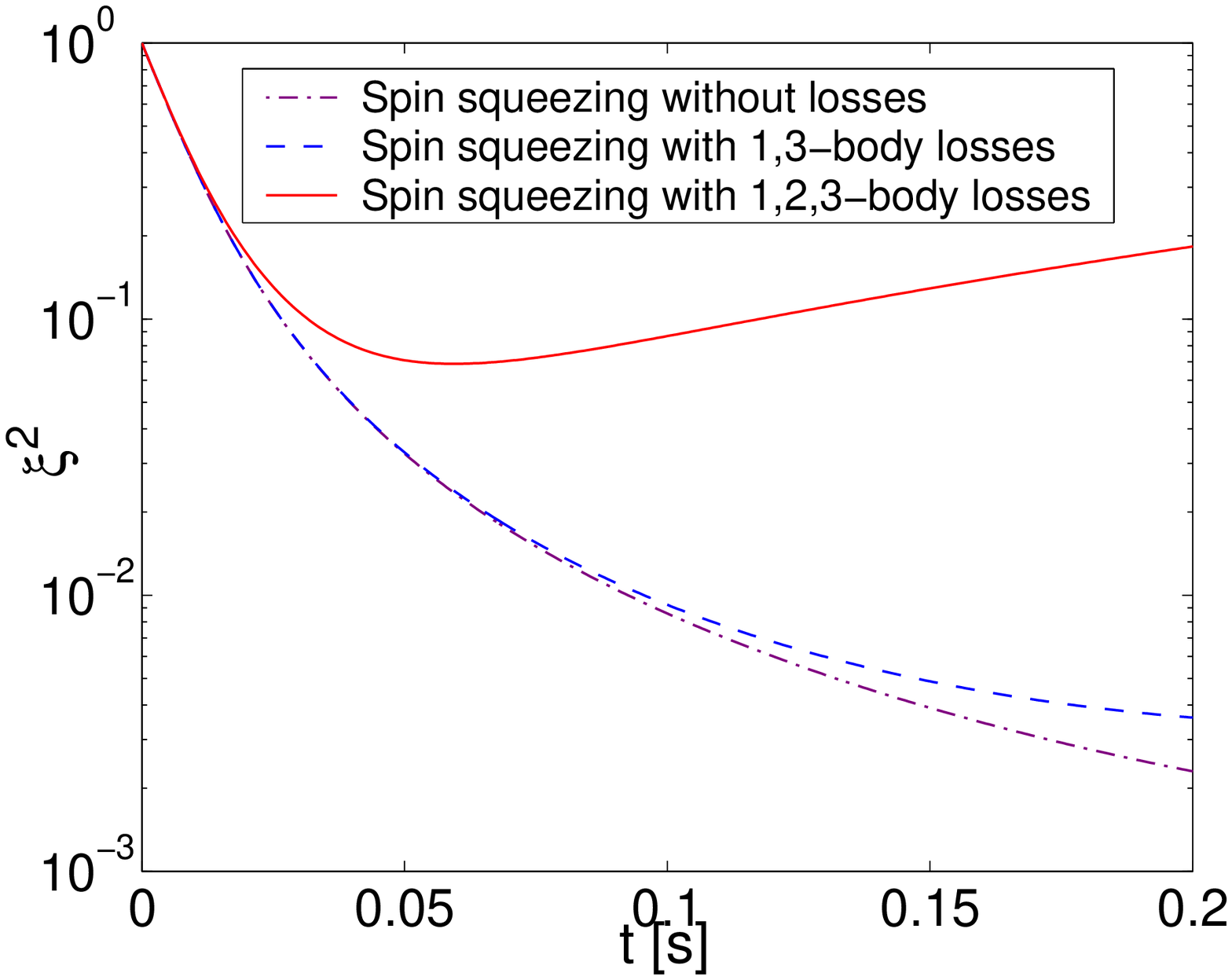}\\
\includegraphics[scale=0.4]{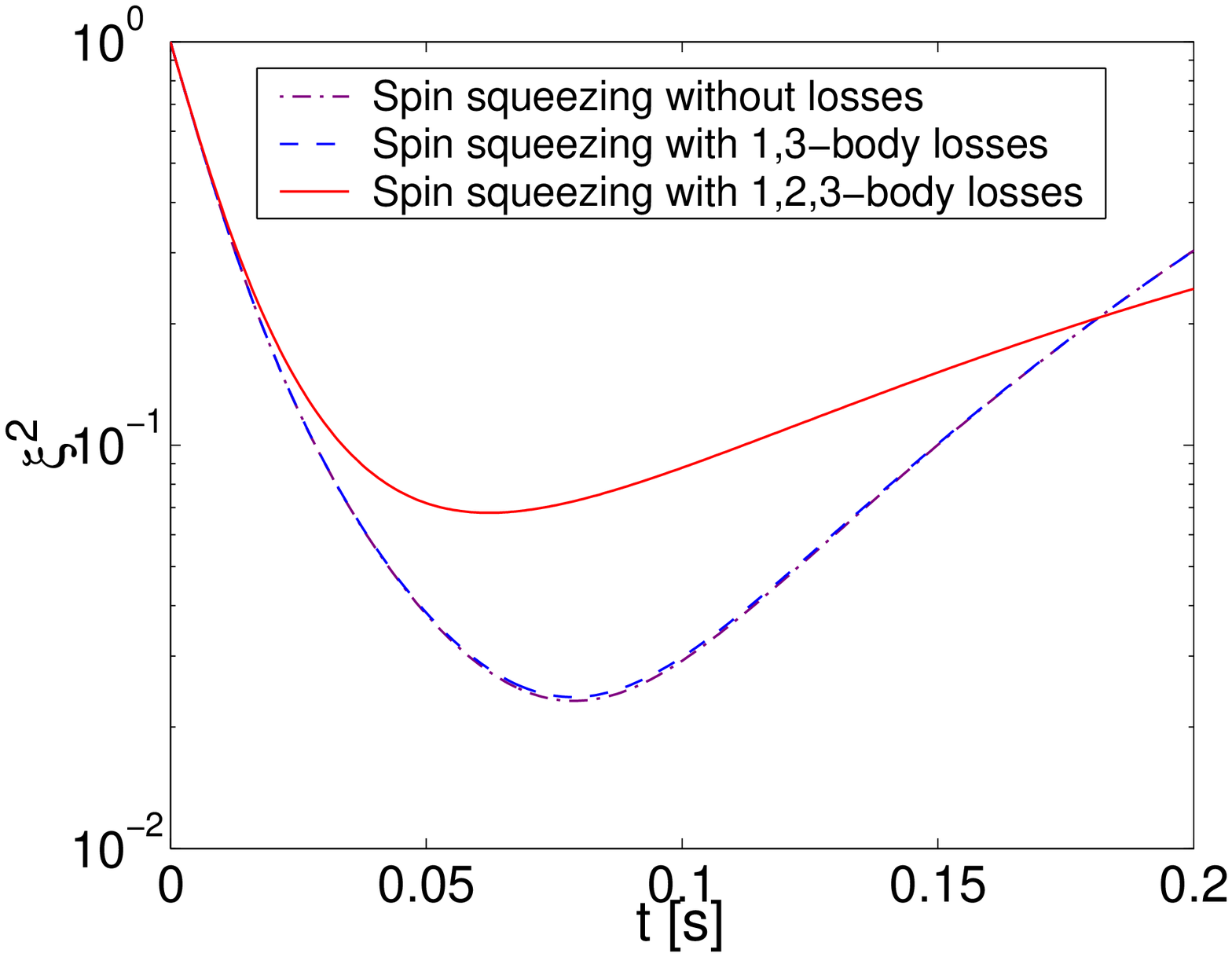}
\caption{Spin squeezing with and without losses in a bimodal Rb
condensate from the stationary two-mode model. A Feshbach resonance
is used to reduce the inter-species scattering length. Violet
dash-dotted line:  without losses. Blue dashed line: with one and
three-body losses. Red solid line: with one, two and three-body
losses. (Top) with a $50-50$ mixing of the two states:
$\bar{N}_a=\bar{N}_b=10^4$, $\chi=5.367\times10^{-3}$s$^{-1}$,
$\tilde{\chi}=5.412\times 10^{-4}$s$^{-1}$. (Bottom) in
breathe-together conditions: $\bar{N}_a=7432$, $\bar{N}_b=12568$,
$\chi=5.392\times10^{-3}$s$^{-1}$,
$\tilde{\chi}=1.386\times10^{-3}$s$^{-1}$. Other parameters:
$\omega=2\pi \times 42.6$ Hz, $m$=87 a.m.u., $a_{aa}=100.44 \, r_B$,
$a_{bb}=95.47 \, r_B$, $a_{ab}=88.28 \, r_B$, $r_B$ is the Bohr
radius. Particle losses: $K_1^{(a)}=K_1^{(b)}=0.01$s$^{-1}$,
$K_2^{(a)}=0$,
$K_2^{(b)}=119\times10^{-21}$m$^3$s$^{-1}$\cite{Hall},
$K_2^{(ab)}=78\times10^{-21}$m$^3$s$^{-1}$\cite{Note_2Bab},
$K_3^{(a)}=6\times10^{-42}$m$^6$s$^{-1}$ \cite{Cornell97}. }
\label{fig:Rboptwithlosses}
\end{figure}

In Fig.\ref{fig:Rboptwithlosses} (Top) and (Bottom) we compare a
situation in which the initial condensate is split evenly in the $a$
and $b$ components to a situation in which the mixing is chosen in
order to satisfy the ``breathe-together'' conditions
(\ref{eq:cond_breath}). For the considered parameters, which are the
same as Fig.\ref{fig:Rbopt} and Fig.\ref{fig:Rboptbre}, the spatial
dynamics is not important and the two-mode model is a good
approximation at all times.

The squeezing in presence of losses is calculated using our general
results of Section \ref{sub:lossasym} for asymmetric condensates.
Although without losses the even splitting is more favorable, with
one, two, three-body losses results are comparable $\xi^2 \simeq 6
\times10^{-2}$. We also show a curve obtained for one and three-body
losses only (dashed-line). It is clear that for the considered Rb
states the dominant contribution for decoherence comes from the
two-body losses in the $F=2$ state severely limiting the maximum
amount of obtainable squeezing.

In the cases considered in Fig.\ref{fig:Rboptwithlosses} asymmetric
two-body losses are very high, we therefore check the validity of
the constant loss rate approximation with an exact Monte Carlo wave
function simulation. The main result is that the constant loss rate
approximation is accurate up to the best squeezing time. A more
complete discussion is presented in Appendix \ref{app:loss_num}.

\subsection{Bimodal BEC of Na atoms}

\begin{figure}
\centering
\includegraphics[scale=0.4]{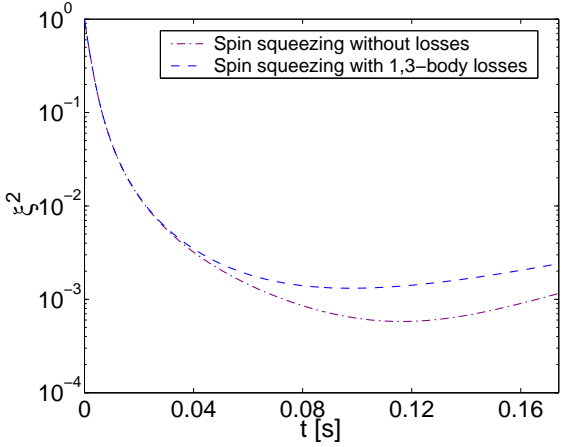}
\caption{Spin squeezing with and without losses in a bimodal Na
condensate from the stationary two-mode model. Violet dash-dotted
line: without losses. Blue dashed line: with one and three-body
losses. Optimized parameters: $\bar{N}_a=\bar{N}_b=4\times 10^{4}$
$\omega=2\pi \times 183$ Hz, $m$=23 a.m.u., $a_{aa}=a_{bb}=51.89 \,
r_B$, $a_{ab}=48.25 \, r_B$, $r_B$ is the Bohr radius.
$\chi=5.517\times10^{-3}$s$^{-1}$, $\tilde{\chi}=0$. Particle
losses: $K_1^{(a)}=K_1^{(b)}=0.01$s$^{-1}$, $K_2^{(a)}=K_2^{(b)}=0$,
$K_3^{(a)}=K_3^{(b)}=2\times10^{-42}$m$^6$s$^{-1}$ \cite{3BNa}. }
\label{fig:Na}
\end{figure}

By using two states in the lower hyperfine manifold, one can greatly
reduce two body losses. A possible example is of using Na atoms in
the $|F=1, m_F=\pm 1\rangle$ states \cite{Nature}. In
Fig.\ref{fig:Na} we calculate the best obtainable squeezing with
these two states. Parameters are chosen according to our
optimization procedure of Section \ref{sub:symm}. A large amount of
squeezing $\xi^2 = 1.9 \times 10^{-3}$ can be reached at the best
squeezing time.

Using our full numerical and our approximated dynamical approaches,
(not shown) we checked that the two-mode model is an excellent
approximation for these parameters.

\section{Dynamically separated Rb BEC}
\label{sec:Galette}

\begin{figure}[htb]
\centering
\includegraphics[scale=0.45]{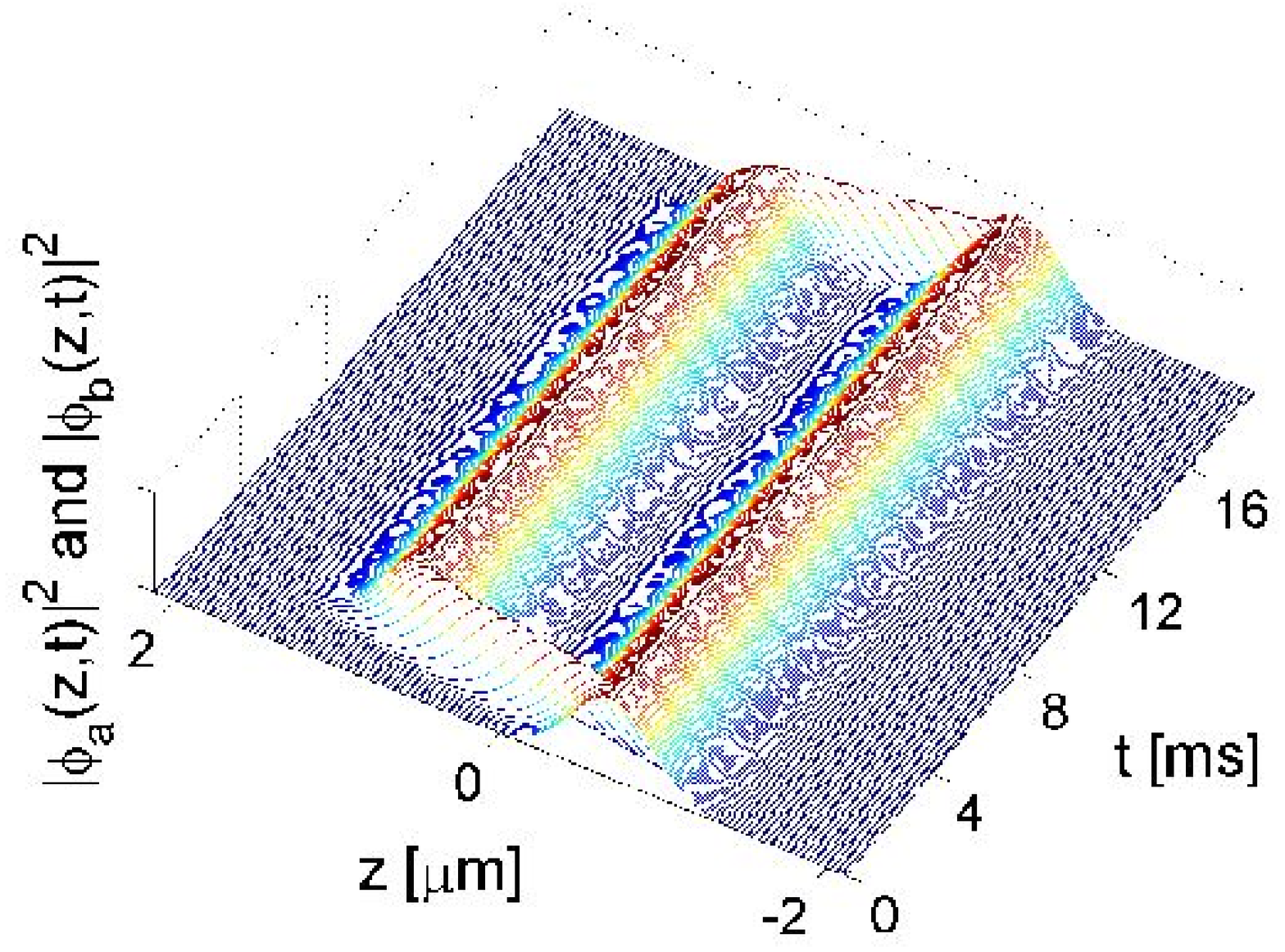} \\
\includegraphics[scale=0.40]{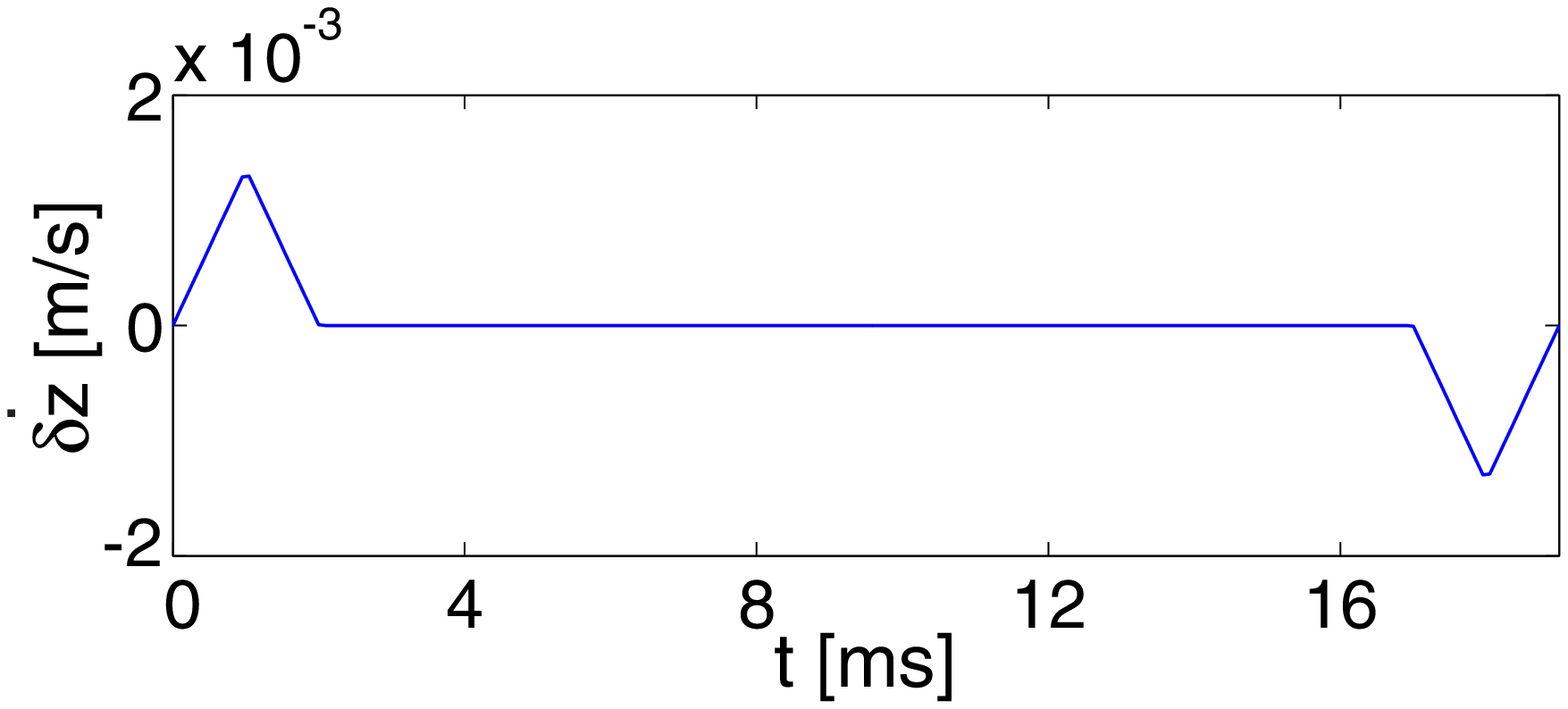}
\caption{(Top) $|\phi_a(z,t)|^2$ and $|\phi_b(z,t)|^2$ in arbitrary
units as the clouds are separated and put back together after an
interaction time of about 15 ms. The harmonic potential for the
$a$-component does not move, while that for the $b$-component is
shifted vertically with a speed $\dot{\delta z}$. The distance
between the two trap centers when they are separated is $\delta z=4
\sqrt{\hbar/M\omega_z}$. (Bottom) variation in time of $\dot{\delta
z}$. Parameters: $\bar{N}_a=\bar{N}_b=5\times 10^4$,
$\omega_{x,y}=2\pi \times 2.31$ Hz, $\omega_{z}=2\pi \times 1$ kHz,
$m$=87 a.m.u., $a_{aa}=100.44 \, r_B$, $a_{bb}=95.47 \, r_B$,
$a_{ab}=98.09 \, r_B$, $r_B$ is the Bohr radius. No particle losses.
} \label{fig:smove}
\end{figure}

In this subsection we consider a bimodal Rb condensate in
$|F=1,m_F=-1\rangle$ and $|F=2,m_F=1\rangle$ states. Rather than
using a Feshbach resonance to change $g_{ab}$, we consider the
possibility of suddenly separating the two clouds right after the
mixing $\pi/2$ pulse using state-selective potentials \cite{Munich},
and recombining them after a well chosen interaction time. A related
scheme using Bragg pulses in the frame of atom interferometry was
proposed in \cite{Uffe}. We consider disc shaped identical traps for
the two states $a$ and $b$ with $\omega_z > \omega_{x,y}\equiv
\omega_\perp$, that can be displaced independently along the $z$
axes. In order to minimize center-of-mass excitation of the cloud,
we use a triangular ramp for the displacement velocity, as shown in
Fig.\ref{fig:smove} (Bottom), with total move-out time
$2\tau=4\pi/\omega_z$ \cite{David}. In Fig.\ref{fig:smove} (Top) we
show the $z$-dependence of densities of the clouds, integrated in
the perpendicular $xy$ plane, as the clouds are separated and put
back together after a given interaction time.

\begin{figure}
\centering
\includegraphics[scale=0.4]{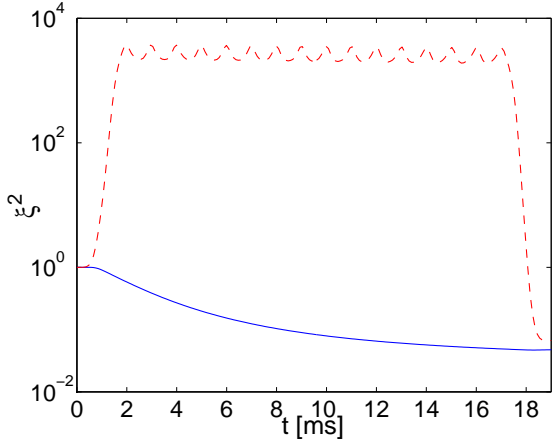}
\caption{Spin squeezing as the two Rb condensates are separated and
put back together after an interaction time of about 15 ms. Red
dashed line:  Spin squeezing obtained from the definitions
(\ref{eq:Sx})-(\ref{eq:Sz}) of the spin operators explicitly
including the overlap between the clouds. Blue solid line: extracted
spin squeezing based of the ``instantaneous modes''
(\ref{eq:extraa})-(\ref{eq:extrbb}). Parameters as in
Fig.\ref{fig:smove}.} \label{fig:Galette}
\end{figure}

We use our dynamical modulus-phase model in 3 dimensions to
calculate the spin squeezing in this scheme.  As the spatial overlap
between the two clouds reduces a lot as they are separated, in
Fig.\ref{fig:Galette} we calculate both the spin squeezing obtained
from the definitions (\ref{eq:Sx})-(\ref{eq:Sz}) of spin operators
(dashed line), and the ``extracted spin squeezing'' introduced in
Section \ref{sub:extract} based on the ``instantaneous modes''
(\ref{eq:extraa})-(\ref{eq:extrbb}) (solid line). The oscillations
in the dashed line are due to tiny residual center of mass
oscillations of the clouds that change periodically the small
overlap between the two modes. They are absent in the extracted spin
squeezing curve (solid line) as they do not affect the spin
dynamics. When the clouds are put back together and the overlap
between the modes is large again, the spin squeezing and the
extracted spin squeezing curves give close results (not identical as
the overlap of the two clouds is not precisely one).

\begin{figure}
\centering
\includegraphics[scale=0.4]{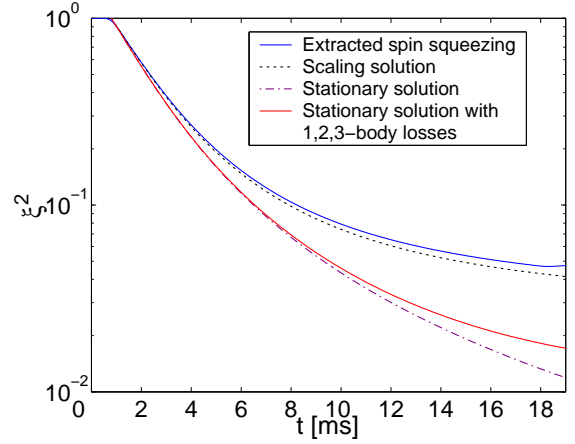}\\
\includegraphics[scale=0.4]{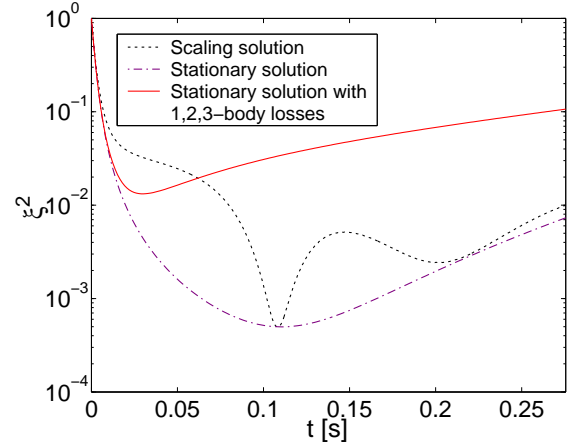}
\caption{Spin squeezing as a function of time. (Top) comparison
between a dynamical calculation and a stationary calculation. Blue
solid line: extracted spin squeezing in 3D. Black doted line: 2D
scaling solution based on (\ref{eq:2Dlambda}). Violet dash-dotted
line: stationary calculation in 3D without losses. Red solid line:
stationary calculation in 3D with losses. Spin squeezing progresses
more slowly in the dynamical calculation than in the stationary
calculation. (Bottom) long time behavior. Black doted line: scaling
solution. Violet dash-dotted line: stationary calculation without
losses. Red solid line: stationary calculation with losses.
Parameters: $\chi=5.003\times10^{-3}$s$^{-1}$,
$\tilde{\chi}=1.342\times10^{-4}$s$^{-1}$,
$K_1^{(a)}=K_1^{(b)}=0.01$s$^{-1}$, $K_2^{(a)}=0$,
$K_2^{(b)}=119\times10^{-21}$m$^3$s$^{-1}$\cite{Hall},
$K_3^{(a)}=6\times10^{-42}$m$^6$s$^{-1}$ \cite{Cornell97}. The other
parameters are as in Fig.\ref{fig:smove}.} \label{fig:GaletteStat}
\end{figure}
In Fig.\ref{fig:GaletteStat} (Top) we compare the extracted spin
squeezing curve of Fig.\ref{fig:Galette} (solid line) with a
two-mode stationary calculation (dash-dotted line) assuming
stationary condensates in separated wells. We notice that the
squeezing progresses much more slowly in the dynamical case. Indeed
when we separate the clouds, the mean field changes suddenly for
each component exciting a breathing mode whose amplitude and
frequency is different for each of the Fock states involved. In the
quasi 2D configuration considered here, the breathing of the wave
functions is well described by a scaling solution in 2D for each
condensate separately \cite{Scaling1,Scaling2} adapted to the case
in which the trap frequency is not changed, but the mean-field is
changed suddenly after separating the two internal states:
\begin{equation}
{\phi}_\varepsilon(\vec{r}_\perp,t)=\frac{e^{-i\eta_\varepsilon(t)}}
{{\cal L}_\varepsilon(t)} e^{i m r_\perp^2 \dot{\cal
L}_\varepsilon(t)/2\hbar{\cal L}_\varepsilon(t)}
{\phi}_0\left(\frac{\vec{r}_\perp}{{\cal L}_\varepsilon(t)\,}\right)
\label{eq:2Dbreath}
\end{equation}
with
\begin{eqnarray}
\dot{\eta}_\varepsilon&=&\frac{\tilde{g}_{\varepsilon \varepsilon}}{\tilde{g}_{aa}} \,
\frac{\bar{\mu}}{{\cal L}_\varepsilon^2\hbar} \\
\frac{d^2{\cal L}_\varepsilon}{dt^2}&=&\frac{N_\varepsilon }{N } \,
\frac{ \tilde{g}_{\varepsilon \varepsilon}}{ \tilde{g}_{aa}}\,
\frac{\omega_\perp^2}{{\cal L}_\varepsilon^3} -\omega_\perp^2 {\cal
L}_\varepsilon \:;
\label{eq:2Dlambda}\\
{\phi}_0(\vec{r}_\perp)&=&\left( \frac{2}{\pi R_0^2}\right)^{1/2}
\left[ 1- \frac{r_\perp^2}{R_0^2}\right]^{1/2} \,.
\end{eqnarray}
$\bar{\mu}$ is the chemical potential of the stationary condensate
{\it before} the $\pi/2$ pulse, when all the $N$ atoms are in state
$a$, $R_0=\sqrt{2\bar{\mu}/m\omega_\perp^2}$ is the corresponding
Thomas-Fermi radius, and $\tilde{g}_{\varepsilon \varepsilon}$ is a
reduced coupling constant to describe the interaction between two
atoms in the $\varepsilon$ condensate in quasi 2D system, where we
assume that the condensate wave functions in the confined direction
are Gaussians:
\begin{equation}
\tilde{g}_{\varepsilon \varepsilon}=\frac{4\pi \hbar^2
a_{\varepsilon \varepsilon}}{M} \sqrt{\frac{M \omega_z}{2 \pi
\hbar}}
\end{equation}
with $a_{\varepsilon \varepsilon}$ the 3D scattering length. The
initial conditions for (\ref{eq:2Dlambda}) are ${\cal
L}_\varepsilon(0)=1$ and $\dot{\cal L}_\varepsilon(0)=0$.

We can use (\ref{eq:2Dbreath}) to calculate the squeezing (dotted
curve) and we note that it reproduces well the spin squeezing curve
obtained integrating 5 Gross-Pitaevskii equations in 3D (full line).
As we studied in detail in Section \ref{sub:breath}, oscillations of
the wave functions cause oscillations of the squeezing parameter due
to entanglement between spatial and spin dynamics. Indeed what we
see in the extracted spin squeezing curve of
Fig.\ref{fig:GaletteStat} (Top) is the beginning of a slow
oscillation for the squeezing parameter. In
Fig.\ref{fig:GaletteStat} (Bottom) we show the long time behavior.
There are indeed times at which the spatial and spin dynamics
disentangle, and the dynamical curve and the steady state curve
touch (see Sect. \ref{sub:breath}). Unfortunately these times are
not accessible here in presence of losses (in particular the high
two-body losses in the higher hyperfine state). Notice that in the
first 15 ms of evolution considered in Fig.\ref{fig:Galette} and
Fig.\ref{fig:GaletteStat} (Top) the effect of losses is small and
the main limitation at short times is provided by the spatial
dynamics.

\begin{figure}
\centering
\includegraphics[scale=0.4]{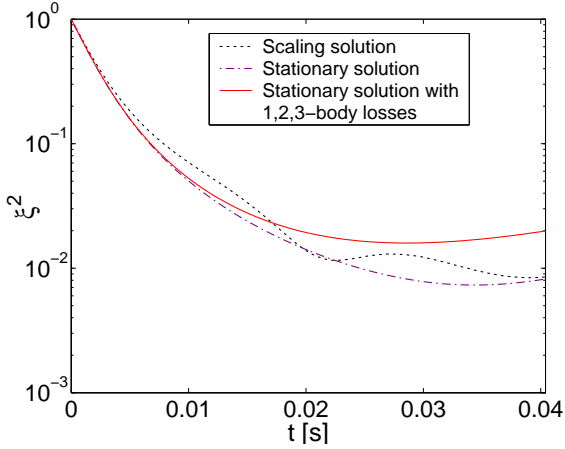}
\caption{Spin squeezing as a function of time in two small Rb
condensates. Black doted line: scaling solution based on
(\ref{eq:2Dlambda}). Violet dash-dotted line: stationary calculation
without losses. Red solid line: stationary calculation with losses.
Parameters: $K_1^{(a)}=K_1^{(b)}=0.01$s$^{-1}$, $K_2^{(a)}=0$,
$K_2^{(b)}=119\times10^{-21}$m$^3$s$^{-1}$\cite{Hall},
$K_3^{(a)}=6\times10^{-42}$m$^6$s$^{-1}$ \cite{Cornell97}. The other
parameters: $\bar{N}_a=\bar{N}_b=10^3$, $\omega_{x,y}=2\pi \times
11.82$ Hz, $\omega_{z}=2\pi \times 2$ kHz, $m$=87 a.m.u.,
$a_{aa}=100.44 \, r_B$, $a_{bb}=95.47 \, r_B$, $a_{ab}=98.09 \,
r_B$, $r_B$ is the Bohr radius. $\chi=0.213$s$^{-1}$,
$\tilde{\chi}=2.763\times10^{-3}$s$^{-1}$. }
\label{fig:Galette1000}
\end{figure}

For a lower number of atoms, the sudden change in the mean field and
the consequent oscillations of the squeezing parameter are reduced.
In Fig.\ref{fig:Galette1000} we show the spin squeezing obtained by
suddenly separating two BEC of Rb atoms in $|F=1,m_F=-1\rangle$ and
$|F=2,m_F=1\rangle$ states with 1000 atoms in each component. The
dotted line is a dynamical calculation using the quasi 2D scaling
solution (\ref{eq:2Dlambda}) (and no losses), while the dash-dotted
line and the solid line are stationary calculations without and with
losses respectively. Note that around $t=0.02$s, where the dynamical
curve and the stationary  curve touch, a squeezing of about $\xi^2
\sim 2 \times 10^{-2}$ could be reached despite the high losses in
the $F=2$ state \cite{NotePhil}.

\section{Conclusions}
In conclusion we developed a method to study the entangled spatial
and spin dynamics in binary mixtures of Bose-Einstein condensates.
The method, which is the natural extension of our work \cite{EPJD}
to the case of spin squeezing, allows a full analytical treatment in
some cases and can be used in the general case to study {\it a
priori} complicated situations in 3D without the need of heavy
numerics. Including the effect of particle losses and spatial
dynamics, we have calculated the maximum squeezing obtainable in a
bimodal condensate of Na atoms in $|F=1,m_F=\pm1 \rangle$ states
when the two condensates overlap in space, and we have calculated
the squeezing in a bimodal Rb condensate in which a Feshbach
resonance is used to reduce the inter-species scattering length as
recently realized experimentally \cite{Widera}. For Rb we also
propose an original scheme in which the two components are spatially
separated using state-dependent potentials, recently realized for
the $|F=1,m_F=-1 \rangle$ and $|F=2,m_F=1 \rangle$ states, and then
recombined after a well chosen squeezing time. With this method we
show that $\xi^2\sim 2\times 10^{-2}$ could be reached in
condensates of 1000 atoms, despite the high two-body losses in the
higher hyperfine state.

\medskip
Yun Li acknowledges support from the ENS-ECNU program, and A.S.
acknowledges stimulating discussions with M. Oberthaler, J. Est\`eve
and K. M{\o}lmer. Our group is a member of IFRAF.

\appendix
\section{Quantum averages of the field operators}
\label{app:averages}

Using equations (\ref{field_statea})-(\ref{field_stateb}), the
averages needed to calculate squeezing parameter can be written in
terms of the wave functions $\phi_a$, $\phi_b$ and the phase factor
$A$ solution of equation (\ref{eq:A(t)}):
\begin{eqnarray}\label{eq:ba}
&&\langle\hat{\psi}_b^\dag(\vec{r})
\hat{\psi}_a(\vec{r})\rangle \nonumber \\
&=&\sum^{N}_{N_a=1}\frac{N!}{(N_a-1)!N_b!}|C_a|^{2(N_a-1)}
|C_b|^{2N_b}C^\ast_b C_a \nonumber \\
&&\times\phi^{\ast}_b(N_a-1,N_b+1,\vec{r})
\phi_a(N_a,N_b,\vec{r})\nonumber \\
&&\times\exp\{i[A(N_a-1,N_b+1)-A(N_a,N_b)]/\hbar\}\nonumber\\
&&\times[\langle \phi_a(N_a-1,N_b+1)|\phi_a(N_a,N_b)\rangle]^{N_a-1}\nonumber \\
&&\times[\langle \phi_b(N_a-1,N_b+1)|\phi_b(N_a,N_b)
\rangle]^{N_b}\,.
\end{eqnarray}
\begin{eqnarray}\label{eq:baab}
&&\langle\hat{\psi}_b^\dag(\vec{r})\hat{\psi}_a^\dag(\vec{r}')
\hat{\psi}_a(\vec{r})\hat{\psi}_b(\vec{r}')\rangle \nonumber\\
&=&\sum^{N-1}_{N_a=1}\frac{N!}{(N_a-1)!(N_b-1)!}|C_a|^{2N_a}
|C_b|^{2N_b}\nonumber \\
&&\times\phi^{\ast}_b(N_a,N_b,\vec{r})\phi^{\ast}_a(N_a,N_b,
\vec{r}')\phi_a(N_a,N_b,\vec{r})\nonumber\\
&&\times\phi_b(N_a,N_b,\vec{r}')\,.
\end{eqnarray}
\begin{eqnarray}\label{eq:bbaa}
&&\langle\hat{\psi}_b^\dag(\vec{r})\hat{\psi}_b^\dag(\vec{r}')
\hat{\psi}_a(\vec{r})\hat{\psi}_a(\vec{r}')\rangle\nonumber\\
&=&\sum^{N}_{N_a=2}\frac{N!}{(N_a-2)!N_b!}|C_a|^{2(N_a-2)}
|C_b|^{2N_b}C^{\ast2}_b C_a^2\nonumber\\
&&\times\phi^{\ast}_b(N_a-2,N_b+2,\vec{r})
\phi^{\ast}_b(N_a-2,N_b+2,\vec{r}')\nonumber\\
&&\times\phi_a(N_a,N_b,\vec{r})\phi_a(N_a,N_b,\vec{r}')\nonumber\\
&&\times\exp\{i[A(N_a-2,N_b+2)-A(N_a,N_b)]/\hbar\}\nonumber\\
&&\times[\langle\phi_a(N_a-2,N_b+2)|\phi_a(N_a,N_b)\rangle]^{N_a-2}\nonumber\\
&&\times[\langle \phi_b(N_a-2,N_b+2)|\phi_b(N_a,N_b)\rangle]^{N_b}\,.
\end{eqnarray}
\begin{eqnarray}\label{eq:bbba}
&&\langle\hat{\psi}_b^\dag(\vec{r})\hat{\psi}_b^\dag(\vec{r}')
\hat{\psi}_b(\vec{r})\hat{\psi}_a(\vec{r}')\rangle\nonumber\\
&=&\sum^{N-1}_{N_a=1}\frac{N!}{(N_a-1)!(N_b-1)!}|C_a|^{2(N_a-1)}
|C_b|^{2N_b}C^\ast_b C_a\nonumber\\
&&\times\phi^{\ast}_b(N_a-1,N_b+1,\vec{r})
\phi^{\ast}_b(N_a-1,N_b+1,\vec{r}')\nonumber\\
&&\times\phi_b(N_a,N_b,\vec{r}) \phi_a(N_a,N_b,\vec{r}')\nonumber\\
&&\times\exp\{i[A(N_a-1,N_b+1)-A(N_a,N_b)]/\hbar\}\nonumber\\
&&\times[\langle \phi_a(N_a-1,N_b+1)|\phi_a(N_a,N_b)
\rangle] ^{N_a-1}\nonumber \\
&&\times[\langle \phi_b(N_a-1,N_b+1)| \phi_b(N_a,N_b)
\rangle]^{N_b-1}\,.
\end{eqnarray}
\begin{eqnarray}\label{eq:aaab}
&&\langle\hat{\psi}_a^\dag(\vec{r})\hat{\psi}_a^\dag(\vec{r}')
\hat{\psi}_a(\vec{r})\hat{\psi}_b(\vec{r}')\rangle\nonumber\\
&=&\sum^{N-1}_{N_a=1}\frac{N!}{(N_a-1)!(N_b-1)!}|C_a|^{2N_a}
|C_b|^{2(N_b-1)}C^\ast_a C_b\nonumber\\
&&\times\phi^{\ast}_a(N_a+1,N_b-1,\vec{r})
\phi^{\ast}_a(N_a+1,N_b-1,\vec{r}')\nonumber\\
&&\times\phi_a(N_a,N_b,\vec{r}) \phi_b(N_a,N_b,\vec{r}')\nonumber\\
&&\times\exp\{i[A(N_a+1,N_b-1)-A(N_a,N_b)]/\hbar\}\nonumber\\
&&\times[\langle \phi_a(N_a+1,N_b-1)|
\phi_a(N_a,N_b)\rangle]^{N_a-1}\nonumber \\
&&\times[\langle \phi_b(N_a+1,N_b-1)| \phi_b(N_a,N_b)
\rangle]^{N_b-1}\,.
\end{eqnarray}
We use these averages to calculate the squeezing in our full
dynamical model. In practice we do not sum over all the Fock states
but over a ``large enough width'' (typically $>6 \sqrt{N}$) around
the average number of atoms $\bar{N}_a$, $\bar{N}_b$. The spin
squeezing is obtained by equation (\ref{eq:xi2gen}) using the
definitions (\ref{eq:Sx})-(\ref{eq:Sz}) for the spin operators.

\section{Quantum averages in the modulus-phase approach}
\label{app:modphase}

Within the modulus-phase approximation, the scalar product of the
wave vectors can be written as
\begin{eqnarray}
&&\langle \phi_a(N_a-\beta,N_b+\beta)|\phi_a(N_a,N_b)\rangle
\nonumber \\
&=&\exp\{i\beta\textstyle\int d^3r|\bar{\phi}_a(\vec{r})|^2
[\chi_0(\vec{r})+\chi_d(\vec{r})]\}\;\;\\
\nonumber\\
&&\langle \phi_b(N_a-\beta,N_b+\beta)|\phi_b(N_a,N_b) \rangle
\nonumber \\
&=&\exp\{i\beta\textstyle\int d^3r |\bar{\phi}_b(\vec{r})|^2
[\chi_0(\vec{r})-\chi_d(\vec{r})]\} \;\;\\
\nonumber \\
&&\langle \phi_b(N_a-\beta,N_b+\beta)|\phi_a(N_a,N_b)\rangle \nonumber\\
&=&\textstyle\int d^3r
\bar{\phi}^\ast_b(\vec{r})\bar{\phi}_a(\vec{r})
\exp[i(N_a-\beta)\chi_d(\vec{r})-iN_b\chi_d(\vec{r})]\nonumber\\
&&\times \exp[i(N-\bar{N})\chi_s(\vec{r}) -i N(|C_a|^2-|C_b|^2)
\chi_d(\vec{r})]\nonumber\\
&&\times\exp[i\beta\chi_0(\vec{r})]\\
\nonumber \\
&&\langle \phi_a(N_a+\beta,N_b-\beta)|\phi_b(N_a,N_b)\rangle\nonumber\\
&=&\textstyle \int d^3r
\bar{\phi}^\ast_a(\vec{r})\bar{\phi}_b(\vec{r})
\exp[-iN_a\chi_d(\vec{r})+i(N_b-\beta)\chi_d(\vec{r})]\nonumber \\
&&\times \exp[-i(N-\bar{N})\chi_s(\vec{r})-i N(|C_a|^2-|C_b|^2)
\chi_d(\vec{r})] \nonumber \\
&&\times\exp[-i\beta\chi_0(\vec{r})]
\end{eqnarray}
where $\beta \in \mathbb{Z}$, and we have used the relation
\begin{eqnarray}
&&\textstyle\int d^3r|\bar{\phi}_\varepsilon|^2
\exp[i(\partial_{N_a}-\partial_{N_b})\theta_\varepsilon
(\bar{N}_a,\bar{N}_b)] \nonumber \\
&&\hspace{8mm}\simeq \exp[\,i\textstyle\int d^3r
|\bar{\phi}_\varepsilon|^2(\partial_{N_a}-\partial_{N_b})
\theta_\varepsilon(\bar{N}_a,\bar{N}_b)]\,.\hspace{5mm}
\end{eqnarray}
By using the
Gross-Pitaevskii equations (\ref{eq:GPE}) for
$\phi_\varepsilon(N_a,N_b)$ and for
$\phi_\varepsilon(\bar{N}_a,\bar{N}_b)$, one obtains
\begin{eqnarray}
&& i\hbar \partial_t \left[ (N_a-\bar{N}_a) \frac{\partial
\theta_\varepsilon}{\partial N_a}+ (N_b-\bar{N}_b) \frac{\partial
\theta_\varepsilon}{\partial N_b}
\right]_{\bar{N}_a,\bar{N}_b} \nonumber \\
&& \hspace{5mm}=(N_\varepsilon - \bar{N}_\varepsilon) g_{\varepsilon
\varepsilon} |\phi_\varepsilon|^2 + (N_{\varepsilon'} -
\bar{N}_{\varepsilon'}) g_{\varepsilon \varepsilon'}
|\phi_{\varepsilon'}|^2  \,, \hspace{8mm}\label{eq:r_uno}
\end{eqnarray}
where $\varepsilon \neq \varepsilon'=a,b$. Using (\ref{eq:r_uno})
together with the initial condition (\ref{ini_cond}), we obtain for
the phase factor $A$ in Eq. (\ref{eq:A(t)})
\begin{eqnarray}
&&[A(N_a-1,N_b+1)-A(N_a,N_b)]/\hbar \nonumber \\
=&&-(N_a-1)\textstyle\int d^3r |\bar{\phi}_a(\vec{r})|^2
[\chi_0(\vec{r})+
\chi_d(\vec{r})]\nonumber\\
&&-N_b\textstyle\int d^3r |\bar{\phi}_b(\vec{r})|^2[\chi_0(\vec{r})
-\chi_d(\vec{r})]\\
\nonumber\\
&&[A(N_a-2,N_b+2)-A(N_a,N_b)]/\hbar\nonumber\\
=&&-2(N_a-2)\textstyle\int d^3r|\bar{\phi}_a(\vec{r})|^2
[\chi_0(\vec{r})+\chi_d(\vec{r})] \nonumber\\
&&-2N_b\textstyle\int d^3r |\bar{\phi}_b(\vec{r})|^2
[\chi_0(\vec{r}) -\chi_d(\vec{r})]\nonumber\\
&&-\textstyle\int d^3r \{|\bar{\phi}_a(\vec{r})|^2
[\chi_0(\vec{r})+\chi_d(\vec{r})]\nonumber\\
&&+|\bar{\phi}_b(\vec{r})|^2 [\chi_0(\vec{r}) -\chi_d(\vec{r})]\}
\end{eqnarray}

The averages and variances of the spin operators equations
(\ref{eq:Sx})-(\ref{eq:Sz}) are obtained by equations
(\ref{eq:ba})-(\ref{eq:aaab}) after spatial integration. We get:
\begin{eqnarray}
&&\textstyle\int d^3r\langle\hat{\psi}_b^\dag(\vec{r})
\hat{\psi}_a(\vec{r})\rangle \nonumber \\
&=& N C^\ast_b C_a \textstyle\int d^3r \bar{\phi}^\ast_b(\vec{r})
\bar{\phi}_a(\vec{r})[|C_a|^2 e^{i\chi_d(\vec{r})}+|C_b|^2
e^{-i\chi_d(\vec{r})}]^{N-1}\nonumber \\
&&\times \exp[i(N-\bar{N}) \chi_s(\vec{r})]
\exp[-i\bar{N}(|C_a|^2-|C_b|^2)
\chi_d(\vec{r})]\nonumber \\
&&\times\exp[i\chi_0(\vec{r})]\,.\\
&&\nonumber \\
&&\textstyle\int d^3rd^3r'\langle\hat{\psi}_b^\dag(\vec{r})
\hat{\psi}_a^\dag(\vec{r}')
\hat{\psi}_a(\vec{r}) \hat{\psi}_b(\vec{r}')\rangle \nonumber\\
&=&N(N-1)|C_a|^2|C_b|^2 \textstyle\int
d^3rd^3r'\bar{\phi}^\ast_b(\vec{r}) \bar{\phi}_a(\vec{r})
\bar{\phi}^\ast_a(\vec{r}') \bar{\phi}_b(\vec{r}')\,.\nonumber\\
&&\\
&&\textstyle\int d^3rd^3r'\langle\hat{\psi}_b^\dag(\vec{r})
\hat{\psi}_b^\dag(\vec{r}')
\hat{\psi}_a(\vec{r})\hat{\psi}_a(\vec{r}')\rangle\nonumber\\
&=&N(N-1)C^{\ast 2}_b C^2_a \textstyle\int d^3rd^3r'
\bar{\phi}^\ast_b(\vec{r})\bar{\phi}_a(\vec{r})
\bar{\phi}^\ast_b(\vec{r}')\bar{\phi}_a(\vec{r}')\nonumber\\
&&\times[|C_a|^2 e^{i\chi_d(\vec{r})+i\chi_d(\vec{r}')}
+|C_b|^2e^{-i\chi_d(\vec{r}) -i\chi_d(\vec{r}')}]^{N-2}\nonumber\\
&&\times\exp\{-i\bar{N}(|C_a|^2-|C_b|^2)
[\chi_d(\vec{r})+\chi_d(\vec{r}')]\}\exp\{2i[\chi_0(\vec{r})\nonumber\\
&&+\chi_0(\vec{r}')]\}\exp\{i(N-\bar{N})[\chi_s(\vec{r})
+\chi_s(\vec{r}')]\}\nonumber\\
&&\times\exp\{-i\textstyle\int
d^3r''(|\bar{\phi}_a|^2[\chi_0(\vec{r}'')
+\chi_d(\vec{r}'')]\nonumber\\
&&+|\bar{\phi}_b|^2[\chi_0(\vec{r}'') -\chi_d(\vec{r}'')])\}\,.\\
&&\nonumber \\
&&\textstyle\int d^3rd^3r'\langle\hat{\psi}_b^\dag(\vec{r})
\hat{\psi}_b^\dag(\vec{r}')
\hat{\psi}_b(\vec{r})\hat{\psi}_a(\vec{r}')\rangle\nonumber\\
&=&N(N-1)C^\ast_b C_a |C_b|^2\textstyle\int d^3r'
\bar{\phi}^\ast_b(\vec{r}')\bar{\phi}_a(\vec{r}')[|C_a|^2
e^{i\chi_d(\vec{r}')}\nonumber\\
&&+|C_b|^2 e^{-i\chi_d(\vec{r}')}]^{N-2}
\exp[-i \bar{N}(|C_a|^2-|C_b|^2)\chi_d(\vec{r}')]\nonumber\\
&&\times\exp[i\chi_0(\vec{r}') -i\chi_d(\vec{r}')]
\exp[i(N-\bar{N})\chi_s(\vec{r}')]\,. \\
&&\nonumber \\
&&\textstyle\int d^3rd^3r'\langle\hat{\psi}_a^\dag(\vec{r})
\hat{\psi}_a^\dag(\vec{r}')
\hat{\psi}_a(\vec{r})\hat{\psi}_b(\vec{r}')\rangle\nonumber\\
&=&N(N-1)C^\ast_a C_b |C_a|^2\textstyle\int d^3r'
\bar{\phi}^\ast_a(\vec{r}')\bar{\phi}_b(\vec{r}')[|C_a|^2
e^{-i\chi_d(\vec{r}')}\nonumber\\
&&+|C_b|^2e^{i\chi_d(\vec{r}')}]^{N-2}
\exp[i \bar{N}(|C_a|^2-|C_b|^2)\chi_d(\vec{r}')]\nonumber\\
&&\times \exp[-i\chi_0(\vec{r}') -i\chi_d(\vec{r}')]
\exp[-i(N-\bar{N})\chi_s(\vec{r}')]\,.
\end{eqnarray}
In the above expressions $\chi_d$, $\chi_s$ and $\chi_0$ are the
space and time dependent functions defined in equations
(\ref{eq:chi_d}), (\ref{eq:chi_s}) and (\ref{eq:chi_0}). In practice
it is sufficient to evolve five wave functions $\phi_a(\vec{r},t)$,
$\phi_b(\vec{r},t)$ for ($\bar{N}_a ,\bar{N}_b \pm \delta N_b$) and
($\bar{N}_a \pm \delta N_a, \bar{N}_b$) with $\delta N_{a,b} \neq 0$
(to calculate numerically $\chi_d$ $\chi_s$ $\chi_0$), and with
$\delta N_{a,b}=0$ (to calculate the central wave functions
$\bar{\phi}_{a,b}$). The spin squeezing is obtained by equation
(\ref{eq:xi2gen}) using the definitions (\ref{eq:Sx})-(\ref{eq:Sz})
for the spin operators.

\section{Equality of $\chi_s$ and $\chi_0$ in the breathe-together configuration}
\label{app:chis_chi0}

Evaluating (\ref{eq:r_uno}) for $\varepsilon=a$, $N_a=\bar{N}_a$;
$\varepsilon=b$, $N_b=\bar{N}_b$ and subtracting the two relations,
on obtains
\begin{equation}
\partial_t \left( \bar{\frac{\partial \theta_a}{\partial N_b}} -
\bar{\frac{\partial \theta_b}{\partial N_a}} \right) = 0
\label{eq:r_due}
\end{equation}
where we used the fact that in breathe-together conditions
$|\bar{\phi}_a|=|\bar{\phi}_b|$. Equation (\ref{eq:r_due}) implies
that the time derivative of $\chi_s-\chi_0$ is zero. As for $t=0$
$\chi_s=\chi_0=0$, we conclude that $\chi_s=\chi_0$ at all times.

\section{Extracted spin squeezing quantum averages}
\label{app:extracted}
By using the instantaneous modes (\ref{eq:extraa})-(\ref{eq:extrbb})
and within the modulus-phase approach, the quantum averages useful
to calculate spin squeezing are expressed in terms of the functions:
\begin{eqnarray}
\chi^{\text{ex}}_d(\vec{r},\vec{r}')&=&\frac{1}{2}(\partial_{N_a}-\partial_{N_b})
[\theta_a(\vec{r})-\theta_b(\vec{r}')](\bar{N}_a,\bar{N}_b) \label{eq:chi_d_ex}\\
\chi^{\text{ex}}_s(\vec{r},\vec{r}')&=&\frac{1}{2}(\partial_{N_a}+\partial_{N_b})
[\theta_a(\vec{r})-\theta_b(\vec{r}')](\bar{N}_a,\bar{N}_b) \label{eq:chi_s_ex}\\
\chi^{\text{ex}}_0(\vec{r},\vec{r}')&=&\frac{1}{2}(\partial_{N_a}-\partial_{N_b})
[\theta_a(\vec{r})+\theta_b(\vec{r}')](\bar{N}_a,\bar{N}_b)\,.\hspace{6mm}\label{eq:chi_0_ex}
\end{eqnarray}
We obtain:
\begin{eqnarray}
\langle\tilde{b}^\dag\tilde{a}\rangle &=& N C^\ast_b C_a
\textstyle\int
d^3r_1 d^3r_2 |\bar{\phi}_b(\vec{r}_1)|^2|\bar{\phi}_a(\vec{r}_2)|^2\nonumber\\
&&\hspace{-8mm}\times[|C_a|^2
e^{i\chi^{\text{ex}}_d(\vec{r}_2,\vec{r}_1)}
+|C_b|^2e^{-i\chi^{\text{ex}}_d(\vec{r}_2,\vec{r}_1)}]^{N-1} \nonumber\\
&&\hspace{-8mm}\times\exp[i(N-\bar{N})\chi^{\text{ex}}_s(\vec{r}_2,\vec{r}_1)
+i\chi^{\text{ex}}_0(\vec{r}_2,\vec{r}_1)]\nonumber \\
&&\hspace{-8mm}\times \exp[-i \bar{N}(|C_a|^2-|C_b|^2) \chi^{\text{ex}}_d(\vec{r}_2,\vec{r}_1)]\\
&&\nonumber \\
\langle\tilde{b}^\dag\tilde{a}^\dag\tilde{a}\tilde{b}\rangle
&=&N(N-1)|C_a|^2|C_b|^2\\
&&\nonumber \\
\langle\tilde{b}^\dag\tilde{b}^\dag\tilde{a}\tilde{a}\rangle
&=&N(N-1)C^{\ast 2}_b C^2_a \textstyle\int d^3r_1 d^3r_2 d^3r_3
d^3r_4\nonumber\\
&&\hspace{-8mm}\times|\bar{\phi}_b(\vec{r}_1)|^2|\bar{\phi}_b(\vec{r}_2)|^2
|\bar{\phi}_a(\vec{r}_3)|^2|\bar{\phi}_a(\vec{r}_4)|^2\nonumber\\
&&\hspace{-8mm}\times\{|C_a|^2
e^{i[\chi^{\text{ex}}_d(\vec{r}_4,\vec{r}_2)
+\chi^{\text{ex}}_d(\vec{r}_3,\vec{r}_1)]}\nonumber\\
&&\hspace{-8mm}+|C_b|^2
e^{-i[\chi^{\text{ex}}_d(\vec{r}_4,\vec{r}_2)
+\chi^{\text{ex}}_d(\vec{r}_3,\vec{r}_1)]}\}^{N-2}\nonumber\\
&&\hspace{-8mm}\times\exp\{2i[\chi^{\text{ex}}_
0(\vec{r}_4,\vec{r}_2)
+\chi^{\text{ex}}_0(\vec{r}_3,\vec{r}_1)]\}\nonumber\\
&&\hspace{-8mm}\times\exp\{-i
\bar{N}(|C_a|^2-|C_b|^2)[\chi^{\text{ex}}_ d(\vec{r}_4,\vec{r}_2)
+\chi^{\text{ex}}_d(\vec{r}_3,\vec{r}_1)]\}\nonumber\\
&&\hspace{-8mm}\times\exp\{i(N-\bar{N})[\chi^{\text{ex}}_s(\vec{r}_4,\vec{r}_2)
+\chi^{\text{ex}}_s(\vec{r}_3,\vec{r}_1)]\}\nonumber\\
&&\hspace{-8mm}\times\exp\{-i\textstyle\int d^3r_5
(|\bar{\phi}_a(\vec{r}_5)|^2[\chi^{\text{ex}}_
0(\vec{r}_5,\vec{r}_5)
+\chi^{\text{ex}}_d(\vec{r}_5,\vec{r}_5)]\nonumber\\
&&\hspace{-8mm}+|\bar{\phi}_b(\vec{r}_5)|^2
[\chi^{\text{ex}}_0(\vec{r}_5,\vec{r}_5)
-\chi^{\text{ex}}_d(\vec{r}_5,\vec{r}_5)])\}
\end{eqnarray}
\begin{eqnarray}
&&\nonumber \\
\langle\tilde{b}^\dag\tilde{b}^\dag\tilde{b}\tilde{a}\rangle
&=&N(N-1)C^\ast_b C_a |C_b|^2\textstyle\int d^3r_1 d^3r_2
|\bar{\phi}_b(\vec{r}_1)|^2|\bar{\phi}_a(\vec{r}_2)|^2 \nonumber\\
&&\hspace{-8mm}\times[|C_a|^2e^{i\chi^{\text{ex}}_d(\vec{r}_2,\vec{r}_1)}
+|C_b|^2e^{-i\chi^{\text{ex}}_d(\vec{r}_2,\vec{r}_1)}]^{N-2} \nonumber \\
&&\hspace{-8mm}\times\exp[i\chi^{\text{ex}}_0(\vec{r}_2,\vec{r}_1)
-i\chi^{\text{ex}}_d(\vec{r}_2,\vec{r}_1)]\nonumber \\
&&\hspace{-8mm}\times\exp[-i \bar{N}(|C_a|^2-|C_b|^2)
\chi^{\text{ex}}_d(\vec{r}_2,\vec{r}_1)]\nonumber\\
&&\hspace{-8mm}\times\exp[i(N-\bar{N})\chi^{\text{ex}}_s(\vec{r}_2,\vec{r}_1)]\\
&&\nonumber \\
\langle\tilde{a}^\dag\tilde{a}^\dag\tilde{a}\tilde{b}\rangle
&=&N(N-1)C^\ast_a C_b |C_a|^2\textstyle\int d^3r_1 d^3r_2
|\bar{\phi}_a(\vec{r}_1)|^2|\bar{\phi}_b(\vec{r}_2)|^2 \nonumber\\
&&\hspace{-8mm}\times[|C_b|^2e^{i\chi^{\text{ex}}_d(\vec{r}_1,\vec{r}_2)}
+|C_a|^2e^{-i\chi^{\text{ex}}_d(\vec{r}_1,\vec{r}_2)}]^{N-2} \nonumber \\
&&\hspace{-8mm}\times\exp[-i\chi^{\text{ex}}_0(\vec{r}_1,\vec{r}_2)
-i\chi^{\text{ex}}_d(\vec{r}_1,\vec{r}_2)]\nonumber \\
&&\hspace{-8mm}\times\exp[i \bar{N}(|C_a|^2-|C_b|^2)
\chi^{\text{ex}}_d(\vec{r}_1,\vec{r}_2)]\nonumber\\
&&\hspace{-8mm}\times\exp[-i(N-\bar{N})\chi^{\text{ex}}_s(\vec{r}_1,\vec{r}_2)]
\end{eqnarray}
In case the wave functions $\bar{\phi}_a$, $\bar{\phi}_b$ are
stationary we recover the stationary two-mode model averages given
in the next appendix in the particular case of no losses. The spin
squeezing is obtained by equation (\ref{eq:xi2gen}) using the
definitions (\ref{spin_internal_x})-(\ref{spin_internal_z}) for the
spin operators.

\section{Quantum averages with one-body losses: Exact solution in
the non symmetric case}
\label{app:1BL}

In this appendix we give the exact result for quantum averages
needed to calculate spin squeezing in the case of a two-mode model
with one-body losses only, in the general non-symmetric case.
\begin{eqnarray}
\langle a^\dag a \rangle&=&|C_a|^2 N \exp(-\gamma_a t)
\phantom{\frac{1}{2}}\\
\langle a^\dag a^\dag a a\rangle&=&|C_a|^4 N(N-1)
\exp(-2\gamma_a t)\phantom{\frac{1}{2}}\\
\langle b^\dag b^\dag b b\rangle&=&|C_b|^4 N(N-1)
\exp(-2\gamma_bt)\phantom{\frac{1}{2}}\\
\langle b^\dag a^\dag a b\rangle&=&|C_b|^2|C_a|^2 N(N-1)
\exp[-(\gamma_a+\gamma_b)t]\phantom{\frac{1}{2}}\\
\langle b^\dag a \rangle&=&C^\ast_b C_b e^{-2ivt} N
\exp\left[-\frac{1}{2}
(\gamma_a+\gamma_b)t\right]L_1^{N-1}\hspace{5mm}\\
\langle b^\dag b^\dag b a\rangle&=&|C_b|^2 C^\ast_b C_a e^{-2ivt}
N(N-1) e^{i\chi t} \nonumber\\
&\times& \exp\left[-\frac{1}{2} (\gamma_a+3\gamma_b)t\right] L_1^{N-2}\\
\langle a^\dag a^\dag a b\rangle&=&|C_a|^2 C^\ast_a C_b e^{2ivt}
N(N-1)  e^{i\chi t}\nonumber\\
&\times&\exp\left[-\frac{1}{2} (3\gamma_a+\gamma_b)t\right]
L_{-1}^{N-2}\\
\langle b^\dag b^\dag a a\rangle&=&C^{\ast2}_b C^2_a e^{-4ivt}
N(N-1)\nonumber\\
&\times& \exp\left[-(\gamma_a+\gamma_b)t\right]L_2^{N-2}
\phantom{\frac{1}{2}}
\end{eqnarray}
where we introduced the function $L_\beta$ with $\beta=-1,1,2$
\begin{eqnarray}
L_\beta&=&\frac{|C_a|^2}{\gamma_a +i\beta(\chi+\tilde{\chi})}
\left[\gamma_a e^{i \beta \tilde{\chi}t}
+i\beta(\chi+\tilde{\chi}) e^{-(\gamma_a+i\beta\chi)t}\right] \nonumber\\
&+&\frac{|C_b|^2}{\gamma_b-i\beta(\chi-\tilde{\chi})} \left[\gamma_b
e^{i \beta \tilde{\chi}t} -i\beta(\chi-\tilde{\chi})
e^{-(\gamma_b-i\beta \chi )t}\right] \nonumber \\
&&
\end{eqnarray}
and $v$ given by (\ref{eq:v}).

\section{Quantum averages with one, two, three-body losses in the
non-symmetric case}
\label{app:twomodes}

In this appendix we give the quantum averages useful to calculate
spin squeezing for the two-mode model in the general non-symmetric
case, in presence of one, two and three-body losses.
\begin{eqnarray}
\langle a^\dag a \rangle&=&|C_a|^2 e^{-\lambda t}
\left[N-(\partial_{\sigma_1}+\partial_{\sigma_2})\right]F_0(\sigma_1,\sigma_2)\\
&&\nonumber \\
\langle a^\dag a^\dag a a\rangle&=&|C_a|^4 e^{-\lambda t}
\left[N-(\partial_{\sigma_1}+\partial_{\sigma_2}-1)\right]\nonumber\\
&\times&\left[N-(\partial_{\sigma_1}+\partial_{\sigma_2})\right]F_0(\sigma_1,\sigma_2)\\
&&\nonumber \\
\langle b^\dag b^\dag b b\rangle&=&|C_b|^4 e^{-\lambda t}
\left[N-(\partial_{\sigma_1}+\partial_{\sigma_2}-1)\right]\nonumber\\
&\times&\left[N-(\partial_{\sigma_1}+\partial_{\sigma_2})\right]F_0(\sigma_1,\sigma_2)\\
&&\nonumber \\
\langle b^\dag a^\dag a b\rangle&=&|C_b|^2|C_a|^2 e^{-\lambda t}
\left[N-(\partial_{\sigma_1}+\partial_{\sigma_2}-1)\right]\nonumber\\
&\times&\left[N-(\partial_{\sigma_1}+\partial_{\sigma_2})\right]F_0(\sigma_1,\sigma_2)\\
&&\nonumber \\
\langle b^\dag a \rangle&=&C^\ast_b C_b e^{-(2iv + \lambda) t}
\left(|C_a|^2e^{-i\chi t}+|C_b|^2e^{i\chi
t}\right)^{N-1} \nonumber\\
&\times&\left[N-(\partial_{\sigma_1}+\partial_{\sigma_2})\right]F_1(\sigma_1,\sigma_2)\\
&&\nonumber \\
\langle b^\dag b^\dag b a\rangle&=&|C_b|^2 C^\ast_b C_a e^{-(2iv +
\lambda) t} \left(|C_a|^2e^{-i\chi t}+|C_b|^2e^{i\chi
t}\right)^{N-2} \nonumber\\
&\times&e^{i\chi t}\left[N-(\partial_{\sigma_1}
+\partial_{\sigma_2}-1) \right]
\left[N-(\partial_{\sigma_1}+\partial_{\sigma_2})\right]\nonumber\\
&\times&F_1(\sigma_1,\sigma_2)\\
&&\nonumber \\
\langle a^\dag a^\dag a b\rangle&=&|C_a|^2 C^\ast_a C_b e^{(2iv -
\lambda) t} \left(|C_b|^2e^{-i\chi t}+|C_a|^2e^{i\chi
t}\right)^{N-2} \nonumber\\
&\times&e^{i\chi t}\left[N-(\partial_{\sigma_1}
+\partial_{\sigma_2}-1)\right]
\left[N-(\partial_{\sigma_1}+\partial_{\sigma_2})\right]\nonumber\\
&\times&G_1(\sigma_1,\sigma_2)\\
&&\nonumber \\
\langle b^\dag b^\dag a a\rangle&=&C^{\ast2}_b C^2_a e^{-(4iv +
\lambda) t} \left(|C_a|^2e^{-2i\chi t}
+|C_b|^2e^{2i\chi t}\right)^{N-2}\nonumber\\
&\times&\left[N-(\partial_{\sigma_1}+\partial_{\sigma_2}-1)\right]
\left[N-(\partial_{\sigma_1}+\partial_{\sigma_2})\right]\nonumber\\
&\times&F_2(\sigma_1,\sigma_2)
\end{eqnarray}
where we introduced the functions $F_\beta(\sigma_1,\sigma_2)$ and
$G_\beta(\sigma_1,\sigma_2)$

\begin{eqnarray}
F_\beta(\sigma_1,\sigma_2)&=& \nonumber \\
&&\hspace{-18mm}\exp\left\{ \sum_{m=1}^3 \frac{ e^{m
\sigma_1}\gamma_a^{(m)} [1-e^{-i m \beta(\chi+\tilde{\chi}) t }]} {i
m \beta(\chi+\tilde{\chi})[|C_a|^2e^{-i  \beta(\chi+\tilde{\chi}) t}
+|C_b|^2e^{i  \beta(\chi-\tilde{\chi}) t}]^m } \right. \nonumber \\
&&\hspace{-18mm} + \left. \frac{e^{m \sigma_2}\gamma_b^{(m)} [e^{i m
\beta(\chi-\tilde{\chi}) t}-1] } {i m
\beta(\chi-\tilde{\chi})[|C_a|^2e^{-i  \beta(\chi+\tilde{\chi}) t}
+|C_b|^2e^{i  \beta(\chi-\tilde{\chi}) t}]^m} \right. \nonumber \\
&&\hspace{-18mm} \left. +\frac{e^{\sigma_1+\sigma_2}\gamma_{ab}
[1-e^{-i 2 \beta \tilde{\chi} t }]} {i 2 \beta \tilde{\chi}
[|C_a|^2e^{-i \beta(\chi+\tilde{\chi}) t}
+|C_b|^2e^{i\beta(\chi-\tilde{\chi}) t}]^2} \right\}
\end{eqnarray}
\begin{eqnarray}
G_\beta(\sigma_1,\sigma_2)&=& \nonumber \\
&&\hspace{-18mm}\exp\left\{ \sum_{m=1}^3 \frac{ e^{m
\sigma_2}\gamma_b^{(m)} [1-e^{-i m \beta(\chi-\tilde{\chi}) t }]} {i
m \beta(\chi-\tilde{\chi})[|C_b|^2e^{-i  \beta(\chi-\tilde{\chi}) t}
+|C_a|^2e^{i \beta(\chi+\tilde{\chi}) t}]^m } \right. \nonumber \\
&&\hspace{-18mm} + \left. \frac{e^{m \sigma_1}\gamma_a^{(m)} [e^{i m
\beta(\chi+\tilde{\chi}) t}-1] } {i m
\beta(\chi+\tilde{\chi})[|C_b|^2e^{-i  \beta(\chi-\tilde{\chi}) t}
+|C_a|^2e^{i  \beta(\chi+\tilde{\chi}) t}]^m} \right. \nonumber \\
&&\hspace{-18mm} \left. -\frac{e^{\sigma_1+\sigma_2}\gamma_{ab}
[1-e^{i 2 \beta \tilde{\chi} t }]} {i 2 \beta \tilde{\chi}
[|C_b|^2e^{-i \beta(\chi-\tilde{\chi}) t}
+|C_a|^2e^{i\beta(\chi+\tilde{\chi}) t}]^2} \right\}
\end{eqnarray}
with $\beta=0,1,2$, and all the expressions should be evaluated in
$\sigma_1=\ln{\bar{N}_a}$, $\sigma_2=\ln{\bar{N}_b}$. The expression
of $v$ is given in (\ref{eq:v}). The spin squeezing is obtained by
equation (\ref{eq:xi2gen}) using the definitions
(\ref{spin_internal_x})-(\ref{spin_internal_z}) for the spin
operators (with $\tilde{a}=a$ and $\tilde{b}=b$).

\section{Test of the constant loss rate approximation for high asymmetric
losses}

\label{app:loss_num} 
\begin{figure}
\centering
\includegraphics[scale=0.4]{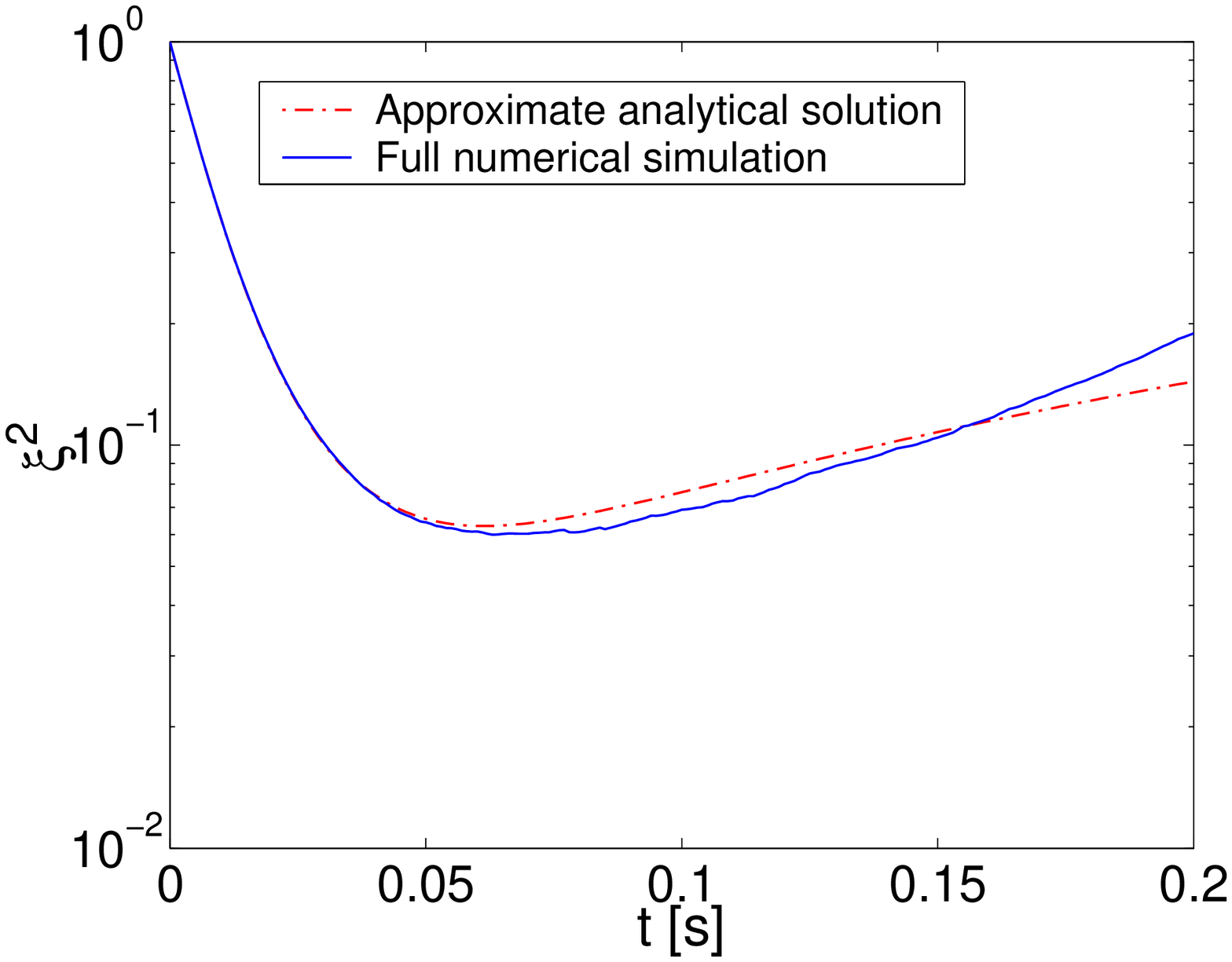}\\
\includegraphics[scale=0.4]{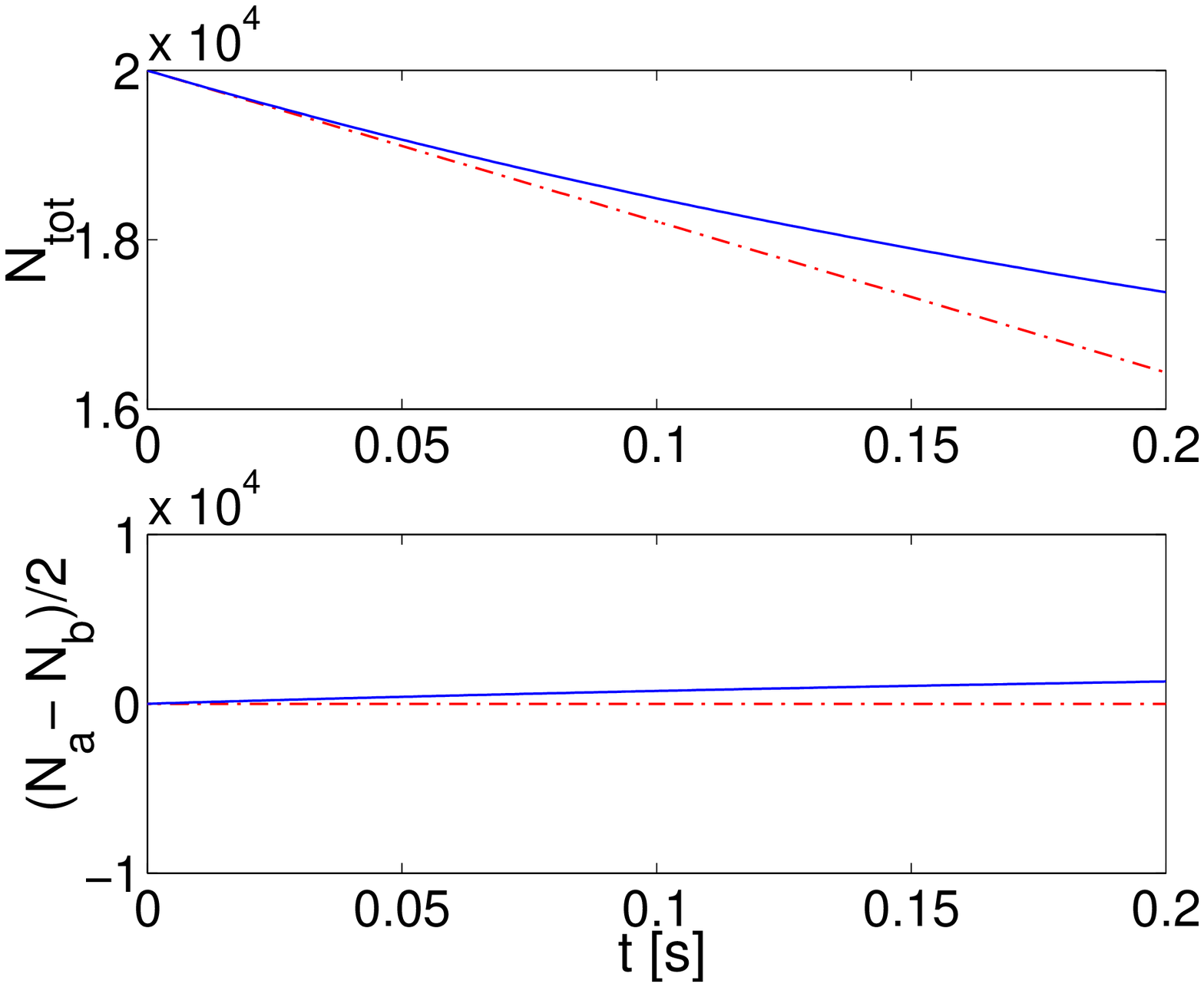}
\caption{(Top) Spin squeezing with two-body losses in a bimodal Rb
condensate as a function of time for symmetrically split
condensates. Blue solid line: exact numerical simulation with 4000
realizations. Red dash-dotted line: analytical solution with
constant loss rate approximation. (Bottom) Corresponding total
number of particles and $\langle S_z \rangle$ as a function of time.
Parameters: $\bar{N}_a=\bar{N}_b=10^4$, $\omega=2\pi \times 42.6$
Hz, $m$=87 a.m.u., $a_{aa}=100.44 \, r_B$, $a_{bb}=95.47 \, r_B$,
$a_{ab}=88.28 \, r_B$, $r_B$ is the Bohr radius,
$\chi=5.367\times10^{-3}$s$^{-1}$, $\tilde{\chi}=5.412\times
10^{-4}$s$^{-1}$, $v=13.758$\,s$^{-1}$. Particle losses:
$K_2^{(a)}=0$, $K_2^{(b)}=119\times10^{-21}$m$^3$s$^{-1}$.}
\label{fig:half_exact}
\end{figure}

\begin{figure}
\centering
\includegraphics[scale=0.4]{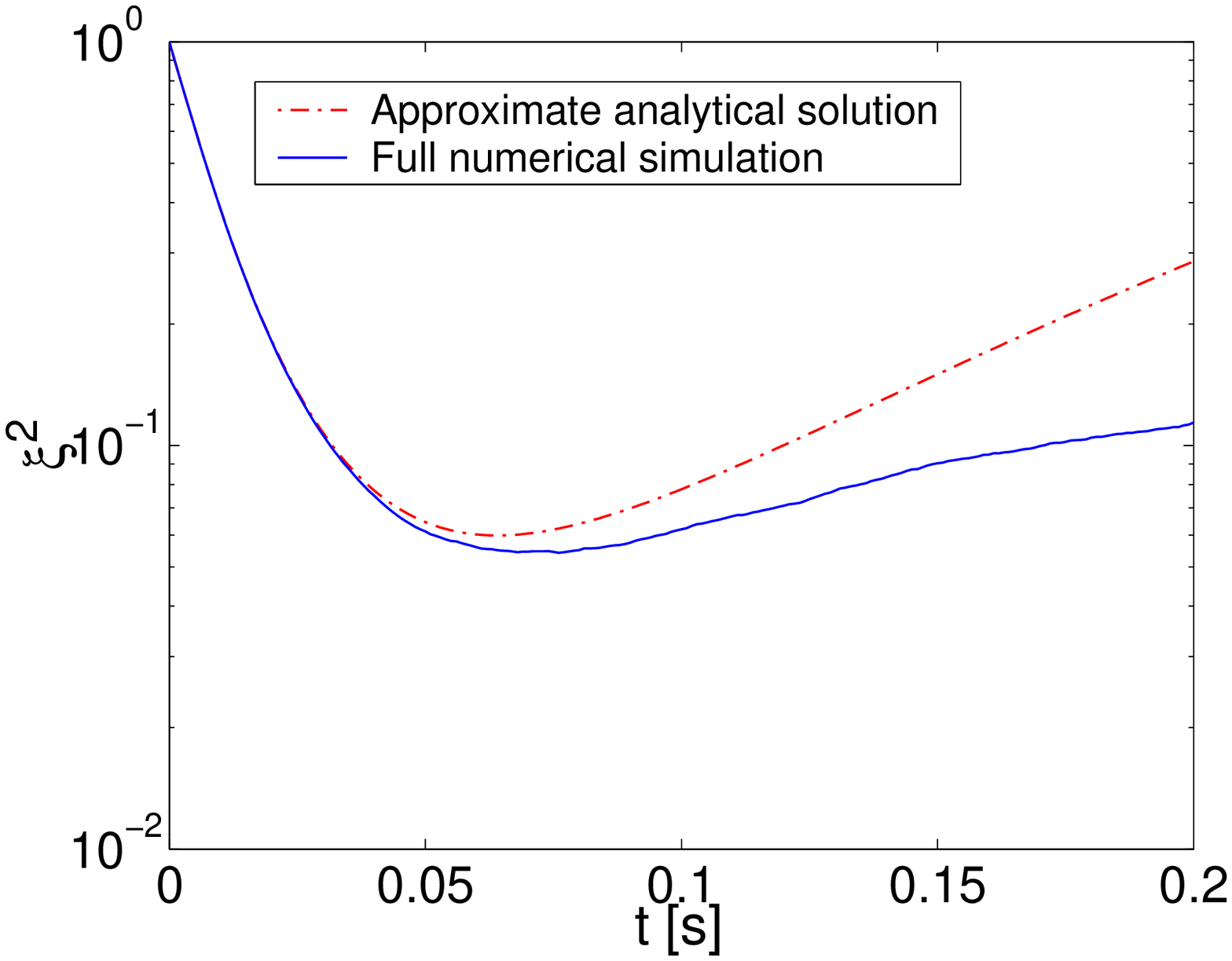}\\
\includegraphics[scale=0.4]{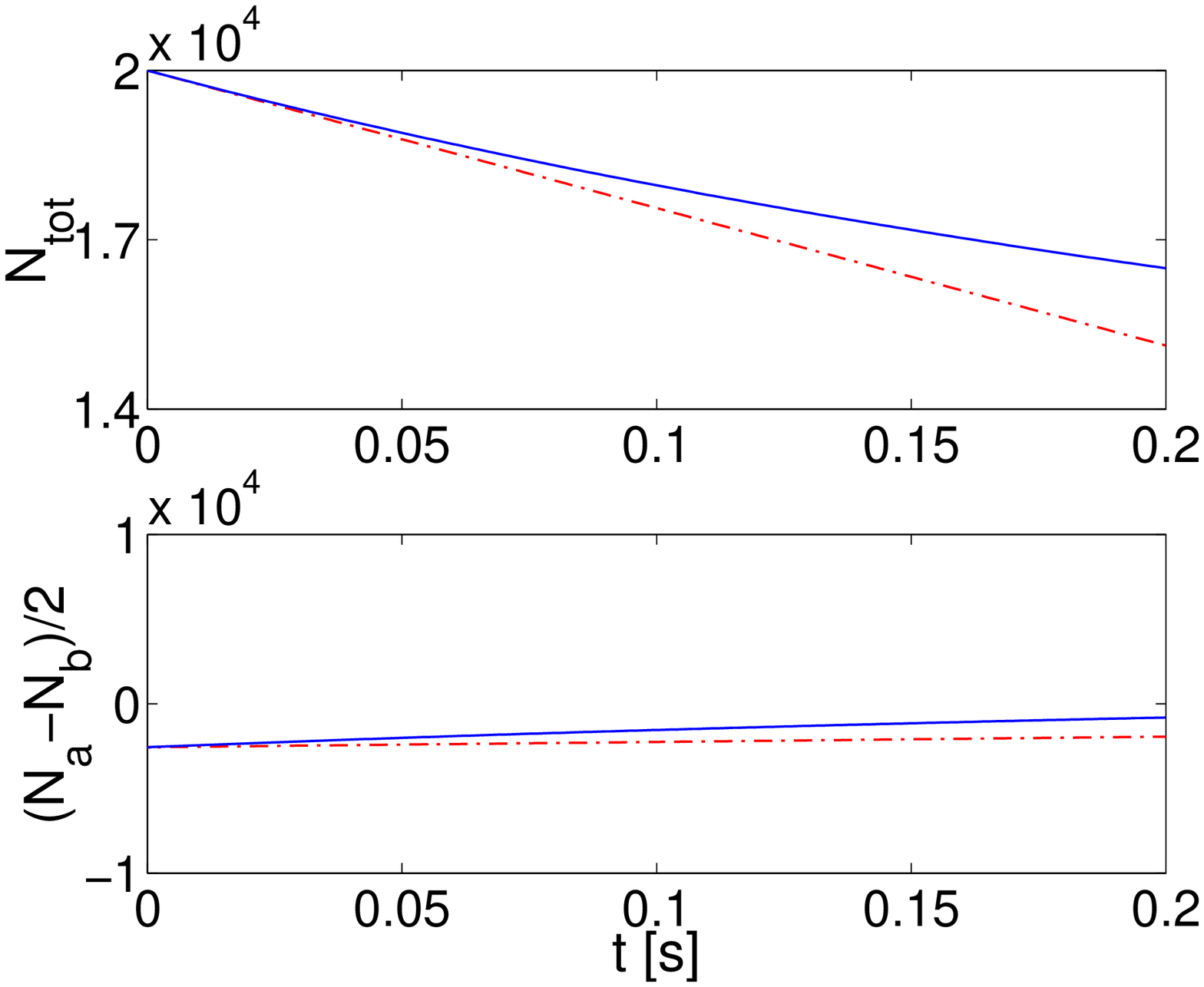}
\caption{(Top) Spin squeezing with two-body losses in a bimodal Rb
condensate as a function of time in breathe-together configuration.
Blue solid line: exact numerical simulation with 4000 realizations.
Red dash-dotted line: analytical solution with constant loss rate
approximation. (Bottom) Corresponding total number of particles and
$\langle S_z \rangle$ as a function of time. Parameters:
$\bar{N}_a=7432$, $\bar{N}_b=12568$,
$\chi=5.392\times10^{-3}$s$^{-1}$,
$\tilde{\chi}=1.386\times10^{-3}$s$^{-1}$, $v=13.850$\,s$^{-1}$. The
other parameters are the same as in Fig.\ref{fig:half_exact}. }
\label{fig:breath_exact}
\end{figure}

The constant loss rate approximation
(\ref{H_eff_app}) is in general valid when a small fraction of
particles is lost. In the case of symmetric condensates, from
equation (\ref{eq:textbook}) one sees that the best squeezing in
presence of losses is of the order of the lost fraction. So that
$\xi(t_{\text{best}}) \ll 1$ guarantees that the lost fraction is
small and the constant loss rate approximation is accurate. In the
case of asymmetric condensates and asymmetric losses there might be
other effects to consider as the population ratio between the two
spin components might change in reality while it remains constant in
the constant loss rate approximation. Indeed with the approximation
(\ref{H_eff_app}), the initial phase state remains a phase state
through out the whole evolution. As a consequence, when a quantum
jump occurs, only the relative phase and the total number of
particle changes (see equation (\ref{eq:jump_on_phasestate})). In
Fig.\ref{fig:half_exact} and Fig.\ref{fig:breath_exact} we compare
the constant loss rate approximation to the exact numerical result
in the case of overlapping Rb condensates with large asymmetric two
body losses considered in Section \ref{sec:results_over}. In
Fig.\ref{fig:half_exact} we address the case of evenly split
condensates $\bar{N}_a=\bar{N}_b=N/2$ while in
Fig.\ref{fig:breath_exact} we address the case of breathe-together
parameters.

The constant loss rate approximation neglects two effects: The
decrease of the loss rate in time as less and less particles are in
the system, and the change of the ratio $\langle N_a \rangle/\langle
N_b \rangle$ as particles from the $b$ component are lost. In the
case of Fig.\ref{fig:half_exact} where we consider initially
$\langle N_a \rangle$=$\langle N_b \rangle$, which is the most
favorable for squeezing, these two effects partially compensates:
one tending to degrade and the other to improve the squeezing with
respect to reality. In the case of Fig.\ref{fig:breath_exact}
instead, the two effects sum-up, both of them tending to degrade the
squeezing with respect to reality. Note however that even for such
large and completely non-symmetric losses, the constant loss rate
approximation proves to be rather accurate up to the best squeezing
time.


\begin{thebibliography}{99}
\bibitem{notecalc1}
We can write the field operators as
$\hat{\psi}_a(\vec{r})={a}_{\langle r|}$,
$\hat{\psi}_b(\vec{r})={b}_{\langle r|}$ and use the commutation
relations:
\begin{eqnarray}\label{field_comm}
&[{a}_{\langle r|},{a}_{|\phi_a(N_a,N_b)\rangle}^\dag ]=
\langle
r|\phi_a(N_a,N_b)\rangle=\phi_a(N_a,N_b,\vec{r})  \nonumber \\
&[\,{b}_{\langle r|},{b}_{|\phi_b(N_a,N_b)\rangle}^\dag\,] =
\langle r|\phi_b(N_a,N_b)\rangle=\phi_b(N_a,N_b,\vec{r}) \,.
\nonumber
\end{eqnarray}
\bibitem{note_alpha}
In the expression (\ref{eq:alpha2}) of $|\alpha_k|^2$ we replaced
$N$ with $\bar{N}$ consistently with the constant loss rate approximation.
\bibitem{Note_2Bab} As we are pretty far from the Feshbach resonance,
we assume for the crossed $ab$ two-body loss rate the same value measured
in \cite{Hall} for the $|F=1,m_F=-1\rangle$, $|F=2,m_F=1\rangle$ states.
\bibitem{NotePhil} We checked that similar result can be obtained with
different geometry where we prepare the condensate in a cigar shape
and separate them along the longitudinal component.
\end{thebibliography}
\end{document}